\documentclass[a4paper,12pt]{article}

\usepackage{amsmath,amssymb,mathtools} 
\usepackage{color}
\usepackage{graphicx}
\usepackage{subfigure}
\usepackage{cite}           
\usepackage{hyperref}            
\usepackage{multirow,makecell}   
\usepackage{textcomp}
\usepackage{wasysym}
\usepackage{setspace}
\usepackage{verbatim}
\usepackage{float}
\usepackage{ulem}
\usepackage[utf8]{inputenc}
\usepackage{soul}

\usepackage[text={17.2cm,25.2cm},centering]{geometry}

\numberwithin{equation}{section}

\begin{document}

\title{\textbf{ Holographic entanglement properties in the QCD phase diagram from EMD models}}
	\author{Zhibin Li $^{1}$\footnote{lizhibin@zzu.edu.cn}}
	\date{}
	
	\maketitle
	
	\vspace{-10mm}
	
	\begin{center}
		{\it
              $^{1}$ Institute for Astrophysics, School of Physics, Zhengzhou University, Zhengzhou 450001, China\\ \vspace{1mm}}
		\vspace{10mm}
	\end{center}

	\begin{abstract}
In this study, we investigate the behavior of entanglement properties in the QCD phase diagram using holographic Einstein-Maxwell-Dilaton (EMD) models. We consider two representative holographic QCD models and examine various entanglement measures, including entanglement entropy, conditional mutual information, and entanglement of purification. We find that the entanglement entropy obtained using the direct UV-cutoff renormalization method shows significant dependence on the specific model being used. However, the outcomes of other non-divergent entanglement measures, such as cutoff-independent entanglement entropy, conditional mutual information, and entanglement of purification, exhibit congruence in our calculations. When we focus specifically on the temperature range below $300$ MeV, we observe a similar behavior among the three entanglement measures. This may suggests that within this temperature range, the entanglement of QCD matter decreases with increasing temperature. Additionally, we observe distinct phase transition behaviors of entanglement properties at the critical end point (CEP) and first-order phase transitions. This observation highlights the potential of entanglement properties as effective order parameters for QCD matter phase transitions.
	\end{abstract}
	
	\baselineskip 18pt
	\thispagestyle{empty}
	\newpage
	
	\tableofcontents

\maketitle
\section{Introduction}\label{sec:01}

The phase structure of quark-gluon matter is a primary focus in heavy-ion collision experiments conducted at the Relativistic Heavy Ion Collider (RHIC) and the Large Hadron Collider (LHC). Numerous efforts have been dedicated to investigating this problem over the past few decades. Lattice QCD calculations suggest that the chiral and confinement/deconfinement phase transitions occur as smooth crossovers at small values of $\mu_B$, with the transitions mixing together~\cite {Borsanyi:2010cj, Borsanyi:2013bia, HotQCD:2014kol}. Conversely, several effective theories, including the Dyson-Schwinger equation (DSE)~\cite{Xin:2014ela, Gao:2016qkh,Qin:2010nq,Shi:2014zpa,Fischer:2014ata,Gao:2020qsj}, the Nambu-Jona-Lasinio (NJL) model~\cite{Asakawa:1989bq,Schwarz:1999dj,Li:2018ygx,Zhuang:2000ub}, and the functional renormalization group (FRG)~\cite{Fu:2019hdw,Zhang:2017icm,Fu:2021oaw}, propose the existence of a first-order phase transition at large values of $\mu_B$. This phase transition would terminate at the QCD critical endpoint (CEP), a critical point that is still subject to debate with no conclusive constraints from model calculations thus far. However, lattice QCD results rule out the presence of the CEP for $\mu_B/T\le 3$ and $\mu_B<300~\text{MeV}$~\cite{Vovchenko:2017gkg,Borsanyi:2020fev,Bazavov:2020bjn,Borsanyi:2021sxv,Bollweg:2022fqq,Philipsen:2021qji}.

In studies investigating the phase transition of strongly coupled quark-gluon matter, thermodynamic quantities and transport properties are commonly employed as order parameters. On the other hand, the study of entanglement properties between different subsystems within QCD matter may offer valuable insights into its strong coupling nature. In quantum information theory, entanglement entropy ($S_{EE}$) is commonly employed to quantify the quantum entanglement between subsystem A and its complement within a pure state. It measures the degree of correlation and information shared between the subsystems due to their entangled state. When dealing with mixed states, the entanglement entropy may not be the most appropriate measure of correlation~\cite{Vidal:2002zz,Plenio:2005cwa,Horodecki:2009zz,Jokela:2020wgs}. Instead, researchers often explore alternatives such as mutual information ($MI$), conditional mutual information ($CMI$), and entanglement of purification ($EoP$). These entanglement properties capture the total correlation between two subsystems and provide finite and scheme-independent results that are always non-negative, a property stemming from the subadditivity and strong subadditivity of $S_{EE}$.

The calculation of entanglement entropy in quantum field theory is complicated because of scheme-dependent behavior at the UV limit. Fortunately, holographic duality provides a clear interpretation of $S_{EE}$: it corresponds to the area of the minimal surface extended in the bulk, with its boundary coinciding with that of subregion $A$~\cite{Ryu:2006bv,Ryu:2006ef,Nishioka:2009un,Rangamani:2016dms,Takayanagi:2017knl}. Holographic QCD~\cite{Zhang:2016rcm,Ali-Akbari:2017vtb,Knaute:2017lll,Dudal:2018ztm,Li:2020pgn,Asadi:2022mvo,Yadav:2022mnv} and holographic condensed matter theory~\cite{Takayanagi:2014rue,Ling:2015dma} research has found that, as non-local observables, entanglement properties can also serve as effective order parameters for various phase transitions. These include transitions such as metal-insulator phase transitions~\cite{Ling:2015dma,Ling:2016dck,Ling:2016wyr}, superconducting phase transitions~\cite{Cai:2012sk,Cai:2012nm,Jeong:2022zea,Yang:2023wuw}, deconfinement phase transitions~\cite{Klebanov:2007ws,Lewkowycz:2012mw,Kol:2014nqa,Jain:2020rbb,Arefeva:2020uec,Jain:2022hxl,daRocha:2021xwq,Chen:2023vjz}, and phase transitions related to quark-gluon matter~\cite{Zhang:2016rcm,Li:2020pgn}. Mutual information~\cite{Mahapatra:2019uql,Ebrahim:2020qif,Walsh:2020lty} and entanglement of purification~\cite{Li:2020pgn,Asadi:2022mvo} have also been utilized as order parameters in several studies characterizing phase transitions.  

Over the past decades, various efforts have been made to apply the gauge/gravity duality to describe the properties of QCD matter, e.g., the properties of mesons \cite{Erlich:2005qh,Karch:2006pv}, baryons \cite{Hong:2007kx,Nawa:2006gv} and glueball \cite{Colangelo:2007pt,Boschi-Filho:2005xct}. Especially, in the soft-wall model, researchers have successfully describe the properties of QCD matter at both zero and finite temperatures by utilizing a quadratic dilaton field, such as the properties of chiral and deconfinement phase transitions \cite{Li:2011hp,Li:2012ay,Fang:2015ytf,Fang:2019lsz}, glueball and baryons \cite{Li:2013oda,Gutsche:2019pls,Gutsche:2019blp}, transport properties \cite{Li:2014dsa} and form factors of hadrons \cite{Gutsche:2017lyu,Gutsche:2019jzh}, etc. On the other hand, the Einstein-Maxwell-Dilaton system which incorporates a bulk nonconformal dilatonic scalar and a $U(1)$ gauge field, has also been extensively utilized in holographic QCD studies \emph{e.g.}~\cite{ DeWolfe:2010he, DeWolfe:2011ts, Cai:2012xh, Cai:2012eh, Finazzo:2013efa, Yang:2014bqa, Critelli:2017oub, Li:2017ple, Chen:2017cyc, Knaute:2017opk, Fang:2018axm, Ballon-Bayona:2020xls, Li:2020hau, Grefa:2021qvt, He:2022amv, Grefa:2022fpu,Chen:2024ckb}. In related studies, it has been observed that the behavior of entanglement properties near phase boundaries can exhibit distinctive characteristics, making it a potentially effective order parameter for phase transitions~\cite{Zhang:2016rcm,Ali-Akbari:2017vtb,Knaute:2017lll,Dudal:2018ztm,Li:2020pgn,Asadi:2022mvo}. However, the behavior of entanglement entropy, even for the same physical process, can vary significantly when different models are utilized. In addition, the dependence of entanglement entropy behavior on the choice of renormalization scheme is another crucial research topic in holographic entanglement entropy. Different renormalization schemes yield distinct behaviors of the entanglement entropy. It has been found that when the entanglement entropy is renormalised in a cut-off independent manner, it is able to reproduce the location of the critical point where a line of first-order phase transitions ends \cite{Jokela:2023lvr}.

In this study, we aim to investigate the behavior of entanglement properties in holographic QCD by adopting two EMD models~\cite{Cai:2022omk,Critelli:2017oub}. These models have been shown to accurately capture the essential characteristics of lattice QCD data with 2+1 flavors~\cite{Borsanyi:2013bia,HotQCD:2012fhj, HotQCD:2014kol, Bazavov:2017dus}. Both models exhibit similar phase structures and possess a critical endpoint (CEP) that aligns with expectations from lattice QCD and heavy ion collision experiments~\cite{Critelli:2017oub,Grefa:2021qvt,Zhao:2022uxc,Li:2023mpv}. In this work, we will examine the behavior of holographic entanglement properties, including the entanglement entropy, conditional mutual information, and entanglement of purification, in the phase diagram. By comparing the behavior of different entanglement properties in different EMD models, we aim to analyze the potential entanglement characteristics near the phase boundary. 

The rest of this paper is organized as follows. In section~\ref{sec:02}, we briefly review the two holographic EMD models. Section~\ref{sec:03} shows the calculation of holographic entanglement entropy and entanglement of purification in standard coordinates. The numerical results for the two holographic QCD models are compared in Section ~\ref{sec:04}. Finally we conclude with a discussion of the findings in Section~\ref{sec:05}.

\section{Review of holographic QCD models}\label{sec:02}
In order to compare and identify common entanglement behaviors among different holographic QCD models within the phase diagram, we consider two distinct models~\cite{Cai:2022omk,Critelli:2017oub} that have been shown to be quantitatively consistent with the equation of state derived from $2+1$ flavor lattice QCD results~\cite{Borsanyi:2013bia,HotQCD:2012fhj, HotQCD:2014kol, Bazavov:2017dus}. Moreover, these models exhibit significant differences in their metric and field configurations, enabling us to analyze the behavior of entanglement properties by examining the results obtained from these models. The gravitational actions of the two models take the same form as
\begin{equation}\label{eq21}
  S_M =\frac{1}{2\kappa_N^2}\int d^5x\sqrt{-g}[\mathcal{R}-\frac{1}{2}\partial_\mu\phi\partial^\mu\phi-\frac{Z(\phi)}{4}F_{\mu\nu}F^{\mu\nu}-V(\phi)]\,.
\end{equation}
Here $\kappa_N^2$ is the bulk gravitational constant, $\mathcal{R}$ is the Ricci scalar and $g_{\mu\nu}$ is the metric of the bulk spacetime. The scalar field $\phi$ is the dilaton, responsible for breaking the
conformal symmetry of the corresponding boundary quantum field theory. Additionally, $F_{\mu\nu}$ denotes the field strength tensor of the vector field $A_\mu$ which incorporating
finite baryon chemical potential and baryon density. Here $V(\phi)$ and $Z(\phi)$ are the dilaton potentials. The bulk fields $\phi$ and $A_\mu$ reads
\begin{equation}\label{eq22}
  \phi=\phi(r),\quad\quad\quad A_\mu dx^\mu=A_t(r)dt,
\end{equation}

\subsection{Holographic QCD Model I}\label{subsec:021}
The metric of the bulk spacetime reads~\cite{Cai:2022omk}
\begin{align}\label{eq23}
d\tilde{s}^2 =-e^{-\tilde{\eta}(\tilde{r})}\tilde{f}(\tilde{r})d\tilde{t}^2+\frac{d\tilde{r}^2}{\tilde{f}(\tilde{r})}+\tilde{r}^2(d \tilde{x}_1^2+d \tilde{x}_2^2+d \tilde{x}_3^2),
\end{align}
with $\tilde{r}$ the holographic radial coordinate and $\tilde{r}\rightarrow \infty$ corresponds to the AdS boundary. Note that here we use superscript $\sim$ to denote this coordinate system as the standard coordinate system, distinguishing it from the numerical coordinate system used for solving the equations of motion. Coordinate transformations between the numerical coordinates and the standard coordinates will be discussed in section \ref{sec:031}. Here $e^{-\tilde{\eta}(\tilde{r})}$ is the wrap factor. We denote the location of event horizon as $\tilde{r}=\tilde{r}_h$ with $\tilde{f}(\tilde{r}_h)=0$. The Hawking temperature and the entropy density are given by
\begin{equation}\label{eq24}
  T=\frac{1}{4\pi}\tilde{f}'(\tilde{r}_h)e^{-\tilde{\eta}(\tilde{r}_h)/2},  \quad  s=\frac{2\pi}{\kappa_N^2}\tilde{r}_h^3.
\end{equation}
Then, by numerically solving the equations of motion, other related thermodynamic quantities, including the energy density $\mathcal{E}$, the pressure $P$, and the trace anomaly $I$ can be obtained using holographic renormalization~\cite{Cai:2022omk}.

In this work we utilize the same couplings configuration of $V(\phi)$ and $Z(\phi)$ presented in ~\cite{Cai:2022omk} 
\begin{equation}\label{eq25}
\begin{split}
 V(\phi)&=-12 \cosh\left(c_1\phi\right)+\left(6c_1^2-\frac{3}{2}\right)\phi^2+c_2\phi^6\,, \\
Z(\phi)&=\frac{1}{1+c_3}\text{sech}(c_4\phi^3)+\frac{c_3}{1+c_3}e^{-c_5\phi}\,,
\end{split}
\end{equation}
which is considered to quantitatively match lattice QCD equation of state data.
In $V(\phi)$ and $Z(\phi)$ parameters $c_1$ to $c_5$ are needed fot the fitting of lattice data. Two additional parameters in the model are the effective Newton constant $\kappa_N^2$ and a characteristic energy scale linked to the source of $\phi$. The latter factor breaks the scale invariance of the boundary system, enabling effective representation of QCD dynamics since real QCD lacks conformal symmetry. These parameters are determined by fitting the lattice QCD data at zero net-baryon density~\cite{HotQCD:2012fhj, HotQCD:2014kol, Bazavov:2017dus} and their specific values can be found in Table~\ref{table1}.
\begin{table}[htbp]
\centering
\begin{tabular}{|c|c|c|c|c|c|c|c|c|}
\hline
    Model I &  $c_1$  &  $c_2$  & $c_3$ & $c_4$  & $c_5$ & $\kappa_N^2$ & $\tilde{\phi}_s $ [MeV] & b \\ \hline
    2+1 flavor  &  0.710  &  0.0037  & 1.935  & 0.091  & 30 & $2\pi(1.68)$  &  1085 & -0.27341 \\ \hline
\end{tabular}
\caption{Parameters for Model I~\cite{Cai:2022omk} by matching the lattice simulation~\cite{HotQCD:2012fhj, HotQCD:2014kol, Bazavov:2017dus}.}
    \label{table1}
\end{table}

\begin{figure}[htbp]
\centering
\includegraphics[width=.45\textwidth]{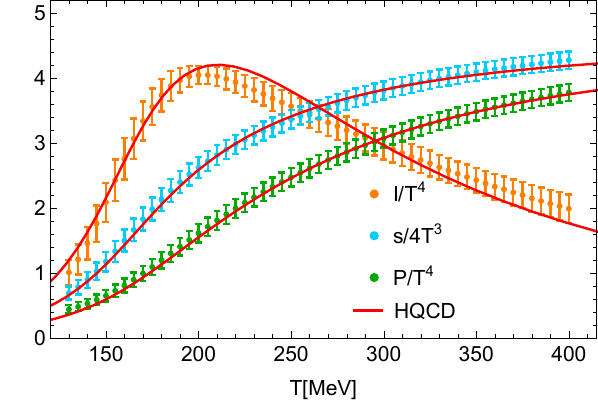}
\qquad
\includegraphics[width=.45\textwidth]{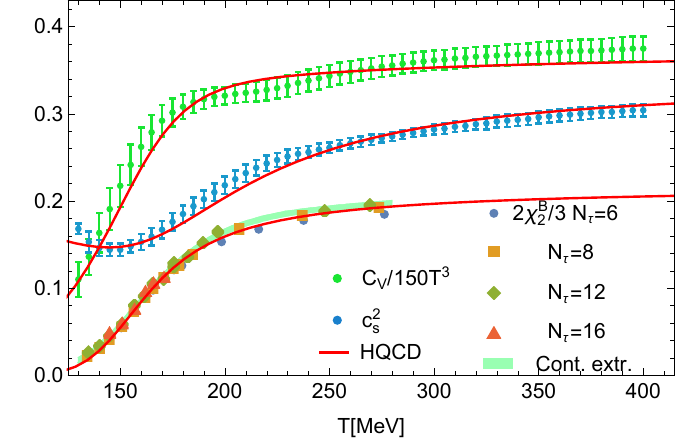}
\caption{Thermodynamics at $\mu_B=0$ from holographic QCD Model I (red solid curves) compared to HotQCD lattice QCD results~\cite{HotQCD:2012fhj, HotQCD:2014kol, Bazavov:2017dus}. \textbf{Left panel}: the entropy density $s$, the pressure $P$, and the trace anomaly $I=\mathcal{E}-3P$. \textbf{Right panel}: the specific heat $C_V$, the squared speed of sound $c_s^2$, and the baryon susceptibility $\chi_B^2$.}\label{eosour}
\end{figure}

The parameter $b$ is related to holographic renormalization and is crucial for matching the lattice QCD simulations at $\mu_B=0$. In Fig.~\ref{eosour}, we present a comparison of different thermodynamic quantities obtained from the holographic setup with lattice simulations. One case demonstrates that the temperature dependence of these quantities exhibits good agreement with lattice QCD results involving $2+1$ flavors~\cite{HotQCD:2012fhj, HotQCD:2014kol, Bazavov:2017dus}.

\subsection{Holographic QCD Model II}\label{subsec:022}

The metric of the bulk spacetime in \cite{Critelli:2017oub} takes the following form
\begin{align}\label{eq26}
d\tilde{s}^2=e^{2\tilde{A}(\tilde{r})}\left[-\tilde{h}(\tilde{r})d\tilde{t}^2+(d \tilde{x}_1^2+d \tilde{x}_2^2+d \tilde{x}_3^2)\right]+\frac{d\tilde{r}^2}{\tilde{h}(\tilde{r})},
\end{align}
where $\tilde{r}$ is the holographic radial coordinate for which $\tilde{r}\rightarrow \infty$ corresponds to the AdS boundary. Similar to section \ref{subsec:021}, here we still use the superscript $\sim$ to denote the metric and coordinates in the standard coordinate system. The location of the event horizon is fixed as $\tilde{r}=\tilde{r}_h$ where $\tilde{f}(\tilde{r}_h)=0$, the Hawking temperature and the entropy density are given by
\begin{equation}\label{eq27}
  T=\frac{e^{\tilde{A}(\tilde{r}_h)}}{4\pi} \left|\tilde{h}'(\tilde{r}_h)\right| \Lambda,  \quad  s=\frac{2\pi}{\kappa_N^2}e^{3\tilde{A}(\tilde{r}_h)}\Lambda^3,
\end{equation}
with $\Lambda$ the characteristic energy scale.
The related thermodynamic quantities, including the energy density $\mathcal{E}$, the pressure $P$, and the trace anomaly $I$ can be obtained using thermodynamic relation (see~\cite{Critelli:2017oub} for more technical details).

\begin{figure}[htbp]
\centering
\includegraphics[width=.465\textwidth]{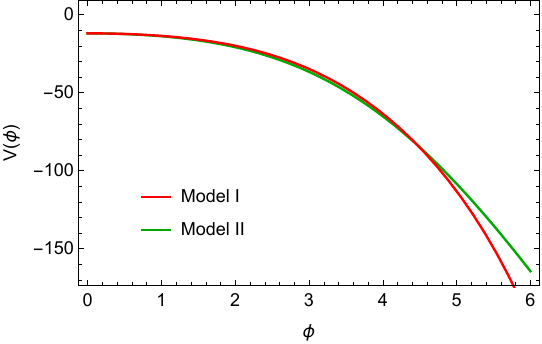}
\qquad
\includegraphics[width=.465\textwidth]{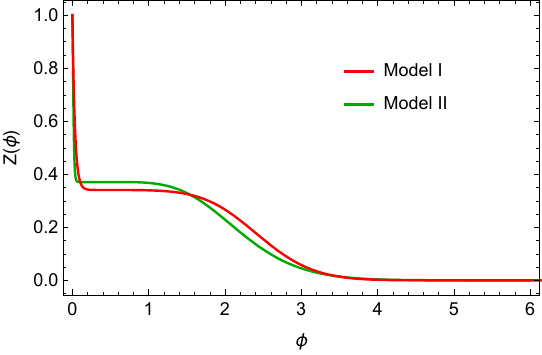}
\caption{Comparison between the dilaton potential $V(\phi)$ and gauge coupling $Z(\phi)$ used in holographic QCD Model I~\cite{Cai:2022omk} and Model II~\cite{Critelli:2017oub}. Both functions obtained by fitting HotQCD and W-B lattice QCD data exhibit certain universal features. Furthermore, the CEP locations differ among different models, with the values of $(T_{CEP}, \mu_B^{CEP})=(105, 555)$ MeV \cite{Cai:2022omk} and $(89, 724)$ MeV \cite{Critelli:2017oub}, respectively. }\label{zvcom}
\end{figure}

\begin{figure}[htbp]
\centering
\includegraphics[width=.465\textwidth]{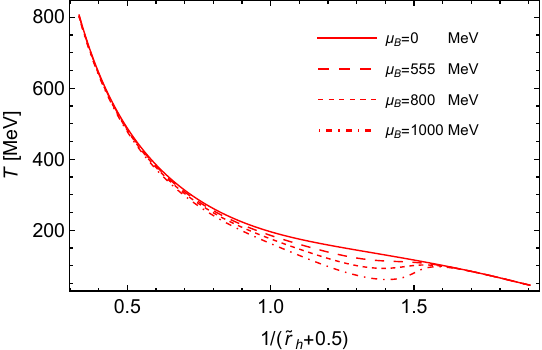}
\qquad
\includegraphics[width=.465\textwidth]{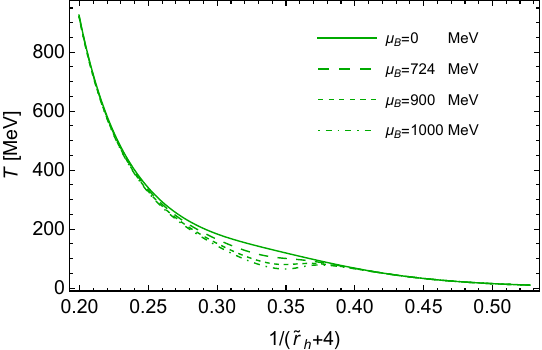}
  \caption{The relationship between black hole horizon $\tilde{r}$ and temperature $T$ from Model I (\textbf{Left panel}) and Model II (\textbf{Right panel}).}\label{trcom}
\end{figure}

The two couplings in~\eqref{eq21} are parameterized as~\cite{Critelli:2017oub}
\begin{equation}\label{eq28}
\begin{split}
 V(\phi)&=-12 \cosh\left(d_1\phi\right)+d_2 \phi^2+d_3 \phi^4+d_4\phi^6\,, \\
Z(\phi)&=\frac{1}{1+d_5}\text{sech}(d_6\phi+d_7\phi^2)+\frac{d_5}{1+d_5}\text{sech}(d_8\phi)\,,
\end{split}
\end{equation}
where $d_1$ to $d_8$ are free parameters that should be fixed by fitting the equation of state of lattice QCD. The other free parameters are the effective Newton constant $\kappa_N^2$ and energy scale $\Lambda$.
All the above parameters are fixed completely by fitting the lattice QCD data at zero net-baryon density~\cite{Borsanyi:2013bia} and their values are summarized in Table.~\ref{table2}.
\begin{table}[htbp]
\centering
\begin{tabular}{|c|c|c|c|c|c|c|c|c|c|c|}
\hline
    Model II &  $d_1$  &  $d_2$  & $d_3$ & $d_4$  & $d_5$ & $d_6$ & $d_7$ & $d_8$ & $\kappa_N^2$ & $\Lambda $ [MeV]  \\ \hline
    2+1 flavor  &  0.63  &  0.65  & -0.05  & 0.003 & 1.7  &  -0.27 & 0.4 & 100 &  $8\pi (0.46)$  &  1058.83   \\ \hline
\end{tabular}
\caption{Parameters of Model II~\cite{Critelli:2017oub} by matching the lattice data~\cite{Borsanyi:2013bia}.}
    \label{table2}
\end{table}
\begin{figure}[htbp]
\centering
\includegraphics[width=.45\textwidth]{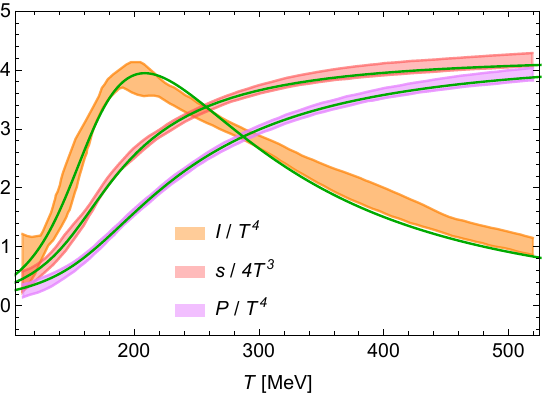}
\qquad
\includegraphics[width=.45\textwidth]{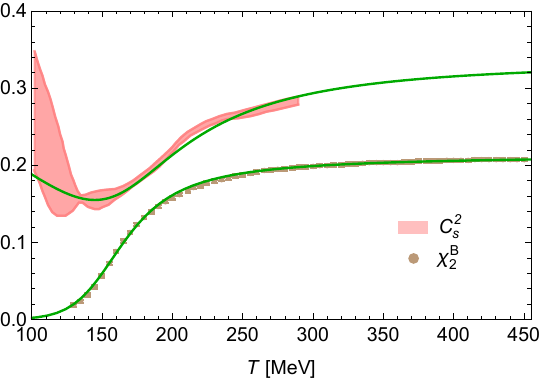}
\caption{Thermodynamics at $\mu_B=0$ from holographic QCD Model II (Green solid curves)~\cite{Critelli:2017oub} compared to lattice QCD results~\cite{Borsanyi:2013bia}. \textbf{Left panel}: the entropy density $s$, the pressure $P$, and the trace anomaly $I=\mathcal{E}-3P$. \textbf{Right panel}: the squared speed of sound $c_s^2$, and the baryon susceptibility $\chi_B^2$.}\label{eosgub}
\end{figure}
Fig.~\ref{eosgub} presents a comparison between the equation of state of lattice QCD~\cite{Borsanyi:2013bia} and the holographic QCD Model II~\cite{Critelli:2017oub}. The behavior of various thermodynamic quantities with temperature is quantitatively matched with the results from lattice QCD.
In Fig.~\ref{zvcom}, a direct comparison of $V(\phi)$ and $Z(\phi)$ functions is shown between Model I and Model II. Notably, both functions, obtained by fitting distinct lattice QCD data, manifest significant universal characteristics. A recent review \cite{Rougemont:2023gfz} also emphasizes a similar comparison and identifies robust features of $V(\phi)$ and $Z(\phi)$ in the EMD description of lattice QCD results with $2+1$ flavors and physical quark masses. Additionally, the location of the CEP (critical endpoint) shows variations among these holographic EMD models, with $(T_{CEP}, \mu_B^{CEP})=(105, 555)$ MeV~\cite{Cai:2022omk} and $(89, 724)$ MeV~\cite{Critelli:2017oub}, respectively.
In Fig.~\ref{trcom}, we observe the relationship between the radius of the black hole horizon $\tilde{r}$ and the temperature $T$. To better illustrate the behavior near the phase boundary, we have chosen $1/(\tilde{r}+0.5)$ and $1/(\tilde{r}+4)$ as the coordinate axes for Model I and Model II, respectively. It can be seen that the relationship between temperature $T$ and horizon radius $\tilde{r}$ in the two models is quite similar. In the crossover region ($\mu_B<\mu_B^{CEP}$), temperature increases monotonically and smoothly as the radius increases. At the CEP ($\mu_B=555~\text{and}~724$ MeV for Model I and Model II, respectively), temperature increases with increasing radius, but there is a plateau where the temperature remains nearly constant while the radius increases. In the case of a first-order phase transition  ($\mu_B>\mu_B^{CEP}$), there will be multiple radii corresponding to a certain range of temperatures.
%
\section{Holographic entanglement properties}\label{sec:03}
\begin{figure}[htbp]
\centering
\includegraphics[width=.6\textwidth]{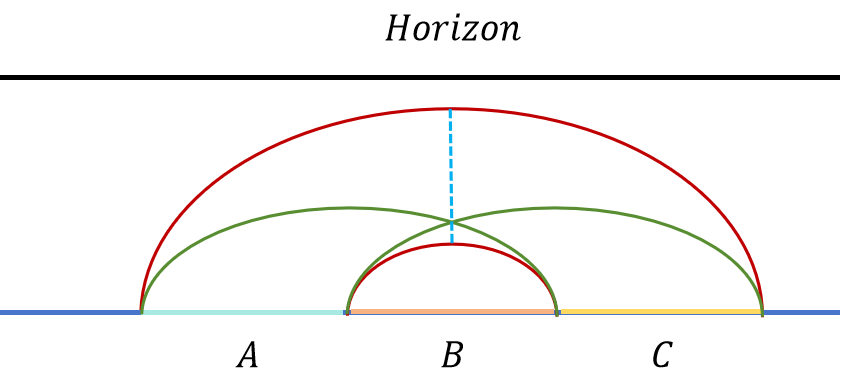}
  \caption{Settings of intervals and their R-T surfaces in this work. We calculate the $S_{EE}$ of one interval with width $\tilde{\ell}$. The conditional mutual information $CMI(A,C|B)$, as shown in equation~\ref{eq33}, can be computed as $2\pi(\text{Area}_{green} - \text{Area}_{red})/\kappa_N^2$. Here, the green lines correspond to the R-T surfaces of the subsystems $A\cup B$ and $B\cup C$, respectively. And the red lines represent the R-T surfaces of $B$ and $A\cup B\cup C$, respectively. Moreover, the minimal surface $\Sigma$ corresponds to the entanglement of purification of $A$ and $C$ is represented by a blue dashed line.}\label{cmieop}
\end{figure}

Entanglement entropy is a measure of the entanglement or quantum correlations between different parts of a quantum system. In field theory, it quantifies the amount of entanglement between different regions of spacetime. The entanglement entropy is defined for a subsystem \emph{e.g.} $B$ within a larger system. It is obtained by tracing out the degrees of freedom in $\bar{B}$ which is the complement of $B$, and then calculating the von Neumann entropy of the reduced density matrix of $B$ as
\begin{equation}\label{eq31}
S_{EE}(B) = -\text{Tr}(\rho_B \ln \rho_B).
\end{equation}
Here, $\rho_B=\text{Tr}_{\bar{B}} \rho$ is the reduced density matrix of subsystem $B$ with $\rho$ the density matrix of the whole system. Computing the entanglement entropy in field theory can be challenging due to the infinite number of degrees of freedom. Following the Ryu-Takayanagi (R-T) formula~\cite{Ryu:2006bv,Ryu:2006ef}, the entanglement entropy is proportional to the area of the R-T surface in the bulk
\begin{equation}\label{eq32}
S_{EE}(B)=\frac{2\pi Area(\Gamma)}{\kappa_N^2},
\end{equation}
with $\Gamma$ the minimal surface in the bulk that is homologous to region $B$ as shown in Fig.~\ref{cmieop}. The red surface in the bulk with the same boundary as $B$ on the boundary represents the R-T surface of $B$. However, as pointed out in~\cite{Vidal:2002zz,Plenio:2005cwa,Horodecki:2009zz,Jokela:2020wgs}, entanglement entropy is not an appropriate measure of entanglement between different subsystems for mixed state. Additionally, other entanglement properties such as the mutual information $MI$ and conditional mutual information $CMI$ are also useful concepts in quantum field theory to quantify the information flow and correlations in quantum field theories. $MI(A,B)$ measures the total correlation between regions $A$ and $B$ and $CMI(A,C|B)$ measures the amount correlations between different subregions $A$ and $C$ conditioned on the measurements in region $B$ which can be calculated as
\begin{equation}\label{eq33}
\begin{split}
MI(A,B)&=S_{EE}(A)+S_{EE}(B)-S_{EE}(AB),\\
CMI(A,C|B)&=S_{EE}(AB)+S_{EE}(BC)-S_{EE}(ABC)-S_{EE}(B)\,.
\end{split}
\end{equation}
Note that $CMI(A,C|B)$ quantifies the remaining correlation between $A$ and $C$ after accounting for the correlation with region $B$ and $CMI(A,C|B)=MI(A,BC)-MI(A,B)=MI(AB,C)-MI(B,C)$. 

Another entanglement property which measures the total correlation between two subsystems $A$ and $C$ in a mixed state is the entanglement of purification $EoP(A,C|B)$. The density matrix of mixed state on $A\cup C$ can be obtained by taking the trace over certain degrees of freedom of the pure-state density matrix as $\rho_{AC}=\text{Tr}_{A^*C^*}\rho_{AA^*CC^*}$. Where $\rho_{AA^*CC^*}$ is a purification of $\rho_{AC}$. Then the entanglement of purification can be defined as
\begin{equation}\label{eq34}
EoP(A,C|B)=\mathop{min}\limits_{\rho_{AC}=\text{Tr}_{A^*C^*}(\rho_{AA^*CC^*})}-\text{Tr}(\rho_{AA^*}\ln \rho_{AA^*}),
\end{equation}
with the reduced density matrix $\rho_{AA^*}=\text{Tr}_{CC^*}(\rho_{AA^*CC^*})$. The minimum in the formula indicates the need to select the purification scheme among various options that minimizes the quantity on the right-hand side of the equation. The holographic correspondence of the entanglement of purification is the minimum of entanglement wedge cross-section (EWCS)
\begin{equation}\label{eq35}
EoP(A,C|B)=\frac{2\pi Area(\Sigma)}{\kappa_N^2}.
\end{equation}
Where $\Sigma$ represents the minimal surface with boundaries on the R-T surfaces of $B$ and $ABC$. In the symmetric case, where the widths of $A$ and $C$ are equal, the minimal surface $\Sigma$ takes the form shown as blue dashed line in Fig.~\ref{cmieop}.

It is important to note that the metrics presented in Section~\ref{sec:02} are given in standard coordinates, which are used for reading thermodynamic quantities. However, in order to numerically solve the equations of motion, one also need to introduce the numerical coordinates that are denoted without a tilde. Additionally, we use the notation $\tilde{\gamma}$ to represent the determinant of the induced metric of the R-T surface, which is a continuous surface in the bulk space with boundaries on the boundary of the bulk space~\cite{Ryu:2006bv,Ryu:2006ef}. Similarly, we denote the determinant of the induced metric of a continuous surface with boundaries on the R-T surfaces as $\tilde{\sigma}$~\cite{Takayanagi:2017knl}. Furthermore, to simplify the calculations, we consider only the symmetric case when evaluating the entanglement of purification. For the calculation of the entanglement entropy, we consider an infinitely long strip region with a fixed width. 

\subsection{HEE in Model I}\label{sec:031}
The numerical coordinates in Model I satisfy the following conditions
\begin{equation}\label{eq36}
   r_h=1,~~~~~~~~~ \eta(r_h)=0.
\end{equation}
While, the standard coordinates satisfy different conditions
\begin{equation}\label{eq37}
   \tilde{\phi}_s=1085~\text{MeV},~~~~~~~~~ \tilde{\eta}(\tilde{r}\rightarrow \infty)=0.
\end{equation}
By setting
\begin{equation}\label{eq38}
    d\tilde{s}^2=ds^2=-e^{-\eta(r)}f(r)dt^2+\frac{dr^2}{f(r)}+r^2(d x_1^2+d x_2^2+d x_3^2)
\end{equation} 
the relation between the standard coordinates and the numerical coordinates can be expressed as follows
\begin{equation}\label{eq39}
\begin{split}
  \tilde{t}=\lambda_t\lambda_r^{-1} t,~~ \tilde{r}=\lambda_r r,~~ \tilde{\eta}(\tilde{r})=\eta(r)& +2\ln{\lambda_t},~~ \tilde{f}(\tilde{r})=\lambda_r^2 f(r), \\
\text{and}~\tilde{x}_i=\lambda_r^{-1}x_i~ & \text{for}~ i=1,~2,~3\,,
\end{split}
\end{equation}
with $\lambda_t=e^{-\eta(\infty)/2}$ and $\lambda_r=1085/\phi_s$.
In Model I the horizon is fixed to $r_h=1$ which correspond to the $\tilde{r}_h=\lambda_r$. The holographic entanglement entropy of Model I in standard coordinates can be calculated as
\begin{equation}\label{eq310}
\begin{split}
  S_{EE}=\frac{2\pi}{\kappa_N^2}\int d\tilde{x}_1 d\tilde{x}_2 d\tilde{x}_3 \sqrt{\tilde{\gamma}}
=&\frac{2\pi \tilde{V}_2}{\kappa_N^2}\int_{-\tilde{\ell}/2}^{\tilde{\ell}/2} d\tilde{x}_1 \tilde{r}^2 \sqrt{\tilde{r}^2+\frac{\tilde{r}'(\tilde{x}_1)^2}{\tilde{f}(\tilde{r})}} \\
=&\frac{2\pi \tilde{V}_2 \lambda_r^2}{\kappa_N^2}\int_{-\ell/2}^{\ell/2} dx_1 r^2 \sqrt{r^2+\frac{r'(x_1)^2}{f(r)}}.
\end{split}
\end{equation}
with $\ell=\lambda_r\tilde{\ell}$. The conserved quantity of integral in~\ref{eq310} shows
\begin{equation}\label{eq311}
  \frac{r^4}{\sqrt{r^2+\frac{r'(x_1)^2}{f(r)}}}=r_{*}^3~~~ \Rightarrow ~~~r'(x_1)=\frac{r}{r_*^3}\sqrt{f(r)(r^6-r_*^6)}
\end{equation}
with $r_*$ the minimum value of $r$ on the R-T surface. The length of the interval shows
\begin{equation}\label{eq312}
\tilde{\ell}=\ell/\lambda_r=\frac{2}{\lambda_r}\int_{r_*}^{\infty}r'(x_1)^{-1}dr=\frac{2}{\lambda_r}\int_{r_*}^{\infty}\frac{r_*^3}{r\sqrt{f(r)(r^6-r_*^6)}}dr.
\end{equation}
Then, the expressions for the holographic entanglement entropy and the entanglement of purification are given as
\begin{equation}\label{eq313}
  S_{EE}=\frac{4\pi \tilde{V}_2 \lambda_r^2}{\kappa_N^2}\int_{r_*}^{\infty}\frac{r^5 }{\sqrt{f(r)(r^6-r_*^6)}}dr,
\end{equation}
and 
\begin{equation}\label{eq314}
  EOP=\frac{2\pi}{\kappa_N^2}\int d\tilde{r} d\tilde{x}_2 d\tilde{x}_3 \sqrt{\tilde{\sigma}}=\frac{2\pi \tilde{V}_2 \lambda_r^2}{\kappa_N^2}\int_{r_{1*}}^{r_{2*}}  \frac{r^2 }{\sqrt{f(r)}} dr.
\end{equation}
%
\subsection{HEE in Model II}\label{sec:032}
The numerical coordinates in Model II satisfy the following conditions
\begin{equation}\label{eq315}
   r_h=0,~~~~~~~~~ h'(r_h)=1.
\end{equation}
On the other hand, the standard coordinates fulfill the following conditions
\begin{equation}\label{eq316}
   \tilde{\phi}(\tilde{r}\rightarrow \infty)=0,~~~~~~~~~ \tilde{h}(\tilde{r}\rightarrow \infty)=1.
\end{equation}
By setting
\begin{equation}\label{eq317}
    d\tilde{s}^2=ds^2=e^{2A(r)}\left[-h(r)dt^2+(d x_1^2+d x_2^2+d x_3^2)\right]+\frac{dr^2}{h(r)}
\end{equation} 
the relation between the standard coordinates and the numerical coordinates in Moedl II can be illustrated as
\begin{equation}\label{eq318}
\begin{split}
  \tilde{r}=\frac{r}{\sqrt{h_0^{\text{far}}}}+A_0^{\text{far}}-\ln{\phi_A^{1/\nu}},~~ \tilde{A}(\tilde{r})=A(r)-& \ln{\phi_A^{1/\nu}} ,~~ \tilde{h}(\tilde{r})=\frac{ h(r)}{h_0^{\text{far}}}, \\
\tilde{t}=\phi_A^{1/\nu} \sqrt{h_0^{\text{far}}} t~\text{and}~\tilde{x}_i=\phi_A^{1/\nu} x_i~ & \text{for}~ i=1,~2,~3\,,
\end{split}
\end{equation}
with $\phi_A$, $A_0^{\text{far}}$ and $h_0^{\text{far}}$ the UV expansion coefficients of field $\phi(r)$, $A(r)$ and $h(r)$ (see~\cite{Critelli:2017oub} for more technical details). 
Model II fix the horizon to $r_h=0$ which means $\tilde{r}_h=A_0^{\text{far}}-\ln{\phi_A^{1/\nu}}$ and the UV boundary at $r=\infty$. 
The holographic entanglement entropy for Model II in standard coordinates can be calculated using the following formula
\begin{equation}\label{eq319}
  S_{EE}=\frac{2\pi \tilde{V}_2}{\kappa_N^2}\int_{-\tilde{\ell}/2}^{\tilde{\ell}/2} d\tilde{x}_1 e^{2 \tilde{A}(\tilde{r})} \sqrt{e^{2 \tilde{A}(\tilde{r})}+\frac{\tilde{r}'(\tilde{x}_1)^2}{\tilde{h}(\tilde{r})}}=\frac{4\pi \tilde{V}_2 \phi_A^{-2/\nu}}{\kappa_N^2}\int_{-\ell/2}^{\ell/2} dx_1 e^{2 A(r)} \sqrt{e^{2 A(r)}+\frac{r'(x_1)^2}{h(r)}},
\end{equation}
with $\ell=\phi_A^{-1/\nu}\tilde{\ell}$ and the conserved quantity
\begin{equation}\label{eq320}
  \frac{e^{4 A(r)}}{\sqrt{e^{2 A(r)}+\frac{ r'(x_1)^2}{h(r)}}}=e^{3 A(r_*)} ~~~\Rightarrow~~~ r'(x_1)=e^{A(r)-3A(r_*)}\sqrt{(e^{6A(r)}-e^{6A(r_*)})h(r)}.
\end{equation}
Then the length of the interval is
\begin{equation}\label{eq321}
\tilde{\ell}=\phi_A^{1/\nu}\ell=2\phi_A^{1/\nu}\int_{r_*}^{\infty}r'(x_1)^{-1}dr=2\phi_A^{1/\nu}\int_{r_*}^{\infty}\frac{e^{3A(r_*)-A(r)}}{\sqrt{(e^{6A(r)}-e^{6A(r_*)})h(r)}}dr.
\end{equation}
And the holographic entanglement entropy and the entanglement of purification in Model II are as
\begin{equation}\label{eq322}
  S_{EE}=\frac{4\pi \tilde{V}_2 \phi_A^{-2/\nu}}{\kappa_N^2}\int_{r_*}^{\infty}\frac{ e^{5A(r)}}{\sqrt{(e^{6A(r)}-e^{6A(r_*)})h(r)}}dr,
\end{equation}
and
\begin{equation}\label{eq323}
  EOP=\frac{2\pi}{\kappa_N^2}\int d\tilde{r} d\tilde{x}_2 d\tilde{x}_3 \sqrt{\tilde{\sigma}}=\frac{2\pi \tilde{V}_2 \phi_A^{-2/\nu}}{\kappa_N^2}\int_{r_{1*}}^{r_{2*}}  \frac{ e^{2A(r)}}{\sqrt{h(r)}} dr.
\end{equation}

In the above formulas \ref{eq310} and \ref{eq319} for entanglement entropy, it is clearly that when $\tilde{r}'(\tilde{x}_1)=0$ and $\tilde{\ell} \rightarrow \infty$, we arrive at the formula for black hole entropy. It implies that in certain scenarios, black hole entropy can be interpreted as a manifestation of entanglement entropy. In fact, it is believed that black hole entropy can be interpreted as a form of entanglement entropy, characterizing the entanglement between the degrees of freedom of field theories on the two boundaries of 5-dimensional bulk spacetime in the Kruskal-Szekeres coordinates~\cite{Ryu:2006bv,Nishioka:2009un,Lewkowycz:2013nqa,Huang:2014aga}.

\section{Numerical results}\label{sec:04}
\begin{figure}[htbp]
\centering
\includegraphics[width=.6\textwidth]{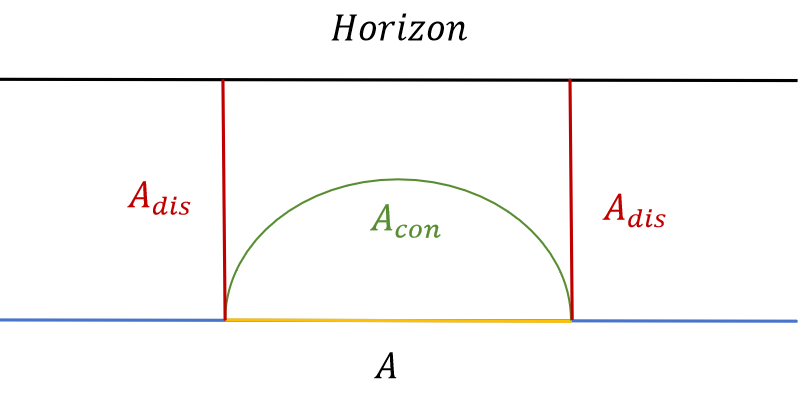}
  \caption{The renormalization of cutoff-independent entanglement entropy $S_{EE}^{ren}$. }\label{seeren}
\end{figure}

In order to handle the divergence in the integral of $S_{EE}$ , two different renormalization schemes were chosen in this work. The renormalized results were denoted as $S_{EE}^{reg}$ and $S_{EE}^{ren}$, respectively. For $S_{EE}^{reg}$, a direct fix of the UV-cutoff was applied. In Model I, the UV-cutoff value was set to $\tilde{r}=200$ . In Model II, the UV-cutoff values were set to $\tilde{r}=4$. In general, directly applying UV-cutoff renormalization to entanglement entropy does not yield precise information about the entanglement properties. However, in theoretical studies, particularly when both the field theory and dual gravity theory are well-defined, and the calculation of entanglement entropy is feasible in both theories, such calculations can offer a more direct test of the gauge/gravity duality~\cite{Ryu:2006bv,Nishioka:2009un,Li:2016pwu,Chen:2017ahf}. In this study, we provide the results of entanglement entropy obtained using direct UV-cutoff renormalization firstly. We also define the cutoff-independent entanglement entropy $S_{EE}^{ren}$ as
\begin{equation}\label{eq328}
  S_{EE}^{ren}=S^{con}_{EE}-S^{dis}_{EE}=\frac{2\pi Area(A_{con})}{\kappa_N^2}-\frac{2\pi Area(A_{dis})}{\kappa_N^2}
\end{equation}
with $A_{con}$ and $A_{dis}$ the area of the connect R-T surfaces (the green surface) and the disconnected surfaces (the red surfaces) as shown in Fig.~\ref{seeren}. We should also fix the 2-dimensional volume term $\tilde{V}=\int_{-\infty}^{\infty}d\tilde{x}_1\int_{-\infty}^{\infty}d\tilde{x}_2$. To ensure consistency and without sacrificing generality, it is often convenient to fix the volume term as a constant. In this work, we fix the volume as $\tilde{V}=\frac{\kappa_N^2}{2\pi}$ in both models. 

The divergence of the entanglement entropy arises from the divergence of the integration at the UV boundary. By employing the R-T formula, it can be demonstrated that the conditional mutual information $CMI(A,C|B)$ is always finite since the divergent part of the entanglement entropy $S_{EE}$ gets completely canceled out. And the value of $CMI(A,C|B)$ is regardless of the specific cutoff and renormalization schemes used for $S_{EE}$. On the other hand, the $MI(A,B)$ is only finite in certain special cases \emph{e.g.} $A$ and $B$ are not jointed. Therefore, we calculate the $CMI$ in this work which provides a more robust measure of information flow and correlations in quantum field theory, as its finiteness is not affected by the details of regularization and renormalization. It is clear that when the width of subregion $B$ is not zero, the entanglement of purification will always be finite, as its integration will not approach the VU boundary. 

\begin{figure}[htbp]
\centering
\includegraphics[width=.465\textwidth]{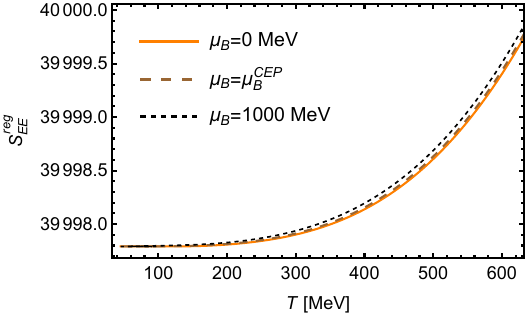}
\qquad
\includegraphics[width=.455\textwidth]{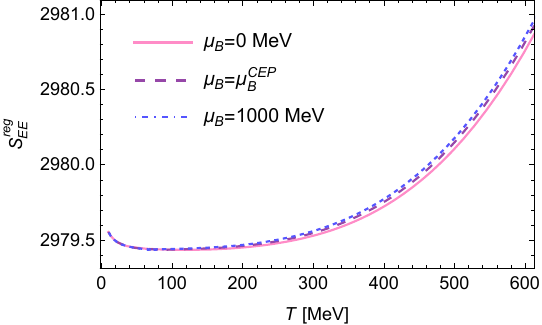}\\
\includegraphics[width=.445\textwidth]{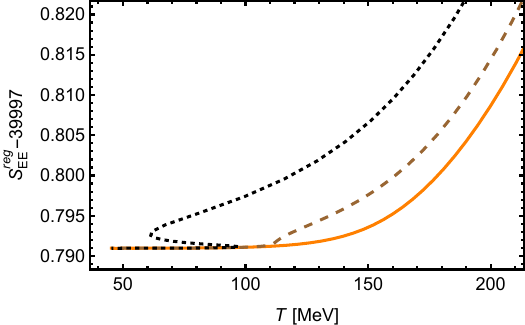}
\qquad
\includegraphics[width=.455\textwidth]{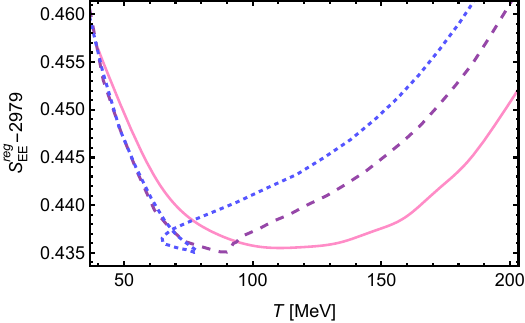}\\
  \caption{The behaviors of entanglement entropy $\ln(S_{EE}^{reg})$ at $\mu_B=0,~\mu_B^{CEP}~\text{and}~1000~\text{MeV}$ from Model I (\textbf{Left panel}) and Model II (\textbf{Right panel}) in standard coordinates.}\label{seephy}
\end{figure}
Fig.~\ref{seephy} illustrates the behavior of the entanglement entropy $\ln(S_{EE}^{reg})$ at $\mu_B=0,\mu_B^{CEP},$ and $1000~\text{MeV}$ for both Model I and Model II in standard coordinates. The top-left and top-right panels showcase the variation of entanglement entropy with temperature in a wide temperature range for Model I and Model II, respectively. The bottom two panels focus on the behavior of entanglement entropy near the phase boundary for both models. It can be observed that over a large temperature range, the behavior of entanglement entropy in both models exhibits a similar trend, with a monotonically increasing pattern in the high-temperature region. However, in the low-temperature region, the entanglement entropy of Model I remains nearly constant, while the entanglement entropy of Model II decreases with increasing temperature, particularly evident in the subsequent two panels. Furthermore, although the entanglement entropy shows smooth, non-smooth monotonic, and multi-valued behavior in the crossover region, CEP, and first-order phase transition region, respectively, there are differences in the behavior of entanglement entropy at the phase boundary between the two models. In particular, the entanglement entropy of Model I exhibits a gradual increase near the phase boundary, while the entanglement entropy of Model II forms a minimum point at the phase boundary. This observation suggests that in standard coordinates, the entanglement entropy $S_{EE}^{reg}$ of the two models exhibits distinct characteristics, with significant qualitative differences in their behavior.

\begin{figure}[htbp]
\centering
\includegraphics[width=.98\textwidth]{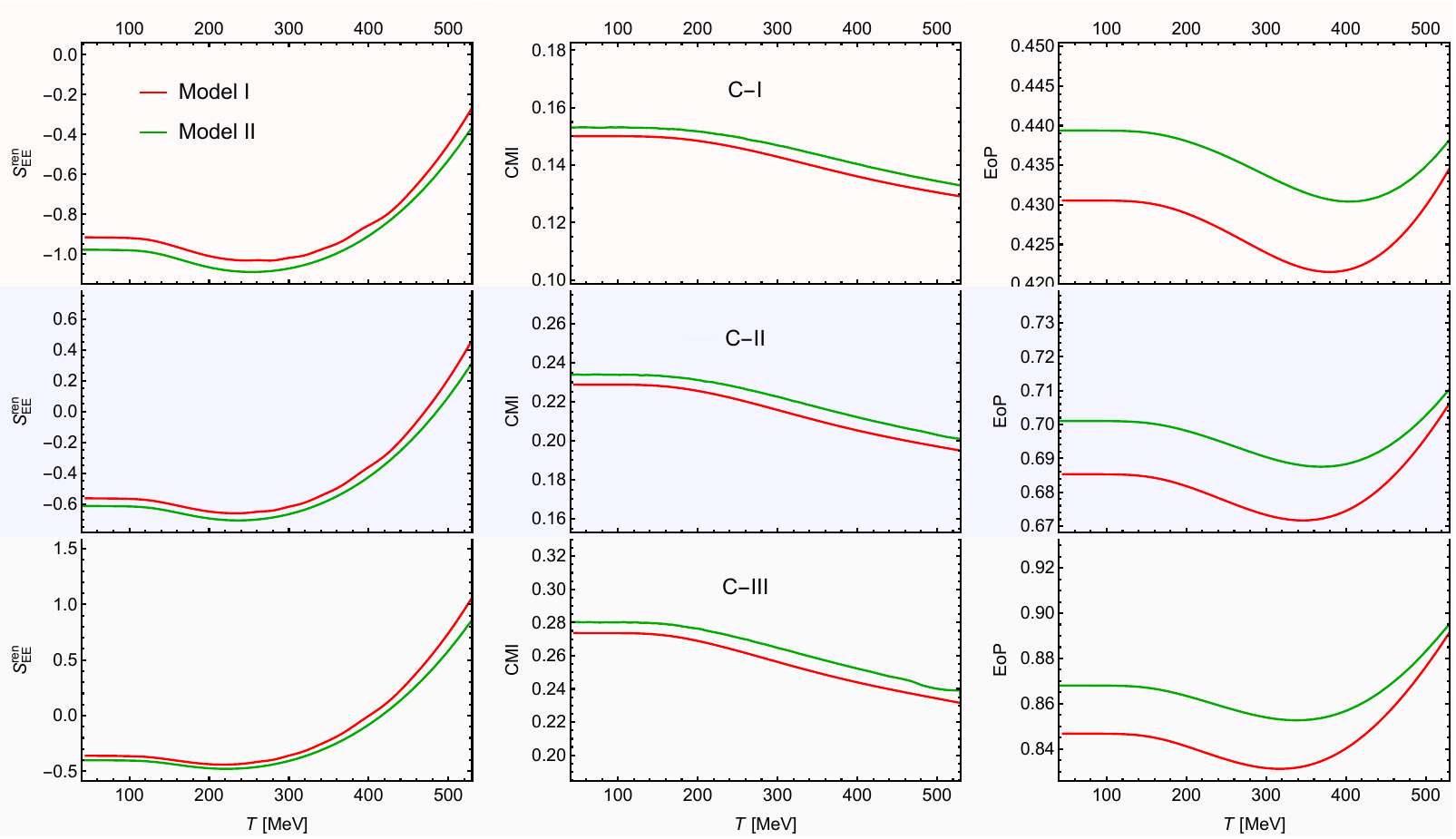}
  \caption{The universal behaviors of entanglement properties $S_{EE}^{ren}$, $CMI$ and $EoP$ for different configurations of intervals in numerical coordinates at $\mu_B=300~\text{MeV}$. The top, middle and bottom three figures are configuration C-I, C-II and C-III as shown in Table~\ref{table4}.}\label{phyfig0}
\end{figure}

Fig.~\ref{phyfig0} presents a comparison of the cutoff-independent entanglement entropy $S_{EE}^{ren}$, conditional mutual information $CMI$, and entanglement of purification $EoP$ in both Model I and Model II. The top, middle, and bottom figures correspond to different interval configurations: C-I, C-II, and C-III, as illustrated in Table~\ref{table4}.
\begin{table}[htbp]
\centering
\begin{tabular}{|c|c|c|c|}
\hline
                 &  C-I  &  C-II  & C-III   \\ \hline
    $S_{EE}^{ren}$  &  $\tilde{\ell}=0.5$  &  $\tilde{\ell}=0.6$  & $\tilde{\ell}=0.7$    \\ \hline
    $CMI$          &  $\tilde{\ell}=0.1, ~0.5, ~0.1$  &  $\tilde{\ell}=0.1, ~0.5, ~0.2$  &  $\tilde{\ell}=0.1, ~0.5, ~0.3$   \\ \hline
    $EoP$    &   $\tilde{\ell}=0.5, ~0.6$  & $\tilde{\ell}=0.6, ~0.7$   &  $\tilde{\ell}=0.7, ~0.8$      \\ \hline
\end{tabular}
\caption{The three different configurations of intervals in standard coordinates for Fig.~\ref{phyfig0}.}
    \label{table4}
\end{table}
We compare the behavior of different entanglement properties at $\mu_B=300~\text{MeV}$ for different interval configurations. The width of the interval for $S_{EE}^{ren}$ is provided, and for $CMI$, the widths of the three intervals corresponding to $A$, $B$, and $C$ in Fig.~\ref{cmieop} are specified. Similarly, the widths of the two intervals corresponding to $B$ and $ABC$ in Fig.~\ref{cmieop} are given for the entanglement of purification. Interestingly, despite the contrasting behavior of $S_{EE}^{reg}$ in Model I and Model II, as depicted in Fig.~\ref{seephy}, the non-divergent entanglements exhibit striking similarities, with negligible quantitative differences. Furthermore, the qualitative behavior of these entanglement quantities remains relatively unchanged across different interval settings, with only quantitative variations.
In the high-temperature regime, both $S_{EE}^{ren}$ and $EoP$ display non-monotonic behavior, characterized by a decrease with increasing temperature at lower temperatures and an increase with increasing temperature at higher temperatures, resulting in the existence of a minimum value. On the other hand, the conditional mutual information exhibits a monotonic decrease across a wide temperature range.

\begin{figure}[htbp]
\centering
\includegraphics[width=.98\textwidth]{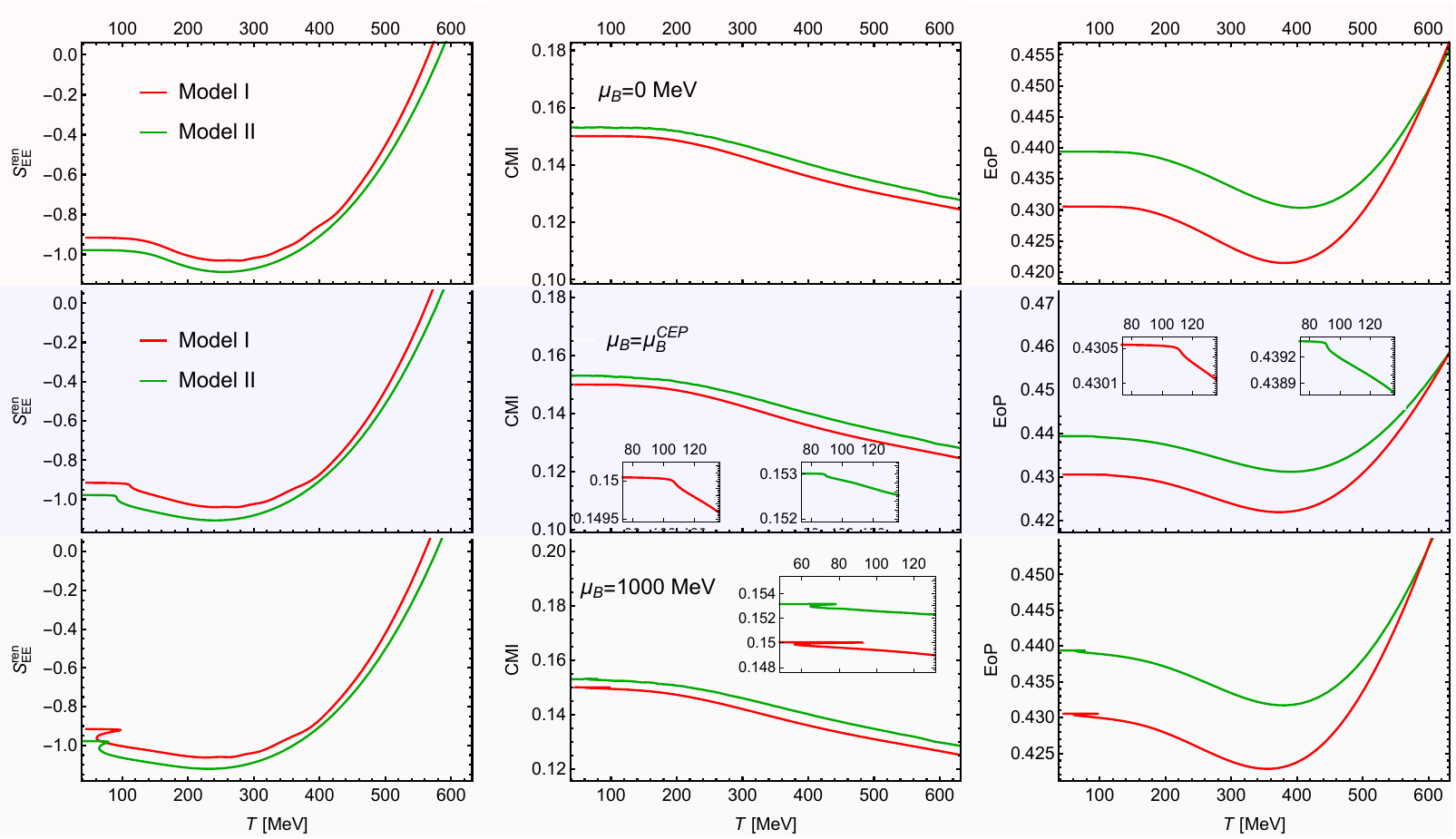}
  \caption{The universal behaviors of entanglement properties $S_{EE}^{ren}$, $CMI$ and $EoP$ at different chemical potential $\mu_B$. The top, middle and bottom three figures are at $\mu_B=0,~\mu_B^{CEP}~\text{and}~1000~\text{MeV}$, respectively.}\label{phyfig}
\end{figure}

To gain insights into how different entanglement properties characterize phase transition behavior, we analyze the behavior of the entanglement properties in the crossover region ($\mu_B=0$), the critical endpoint (CEP) ($\mu_B=\mu_B^{CEP}$), and the first-order phase transition region ($\mu_B=1000~\text{MeV}$) in Fig.~\ref{phyfig}. We present the behavior of $S_{EE}^{ren}$, $CMI$, and $EoP$ at fixed chemical potentials $\mu_B=0,\mu_B^{CEP},$ and $1000~\text{MeV}$, respectively. We observe different characteristics for different chemical potential: smooth behavior, continuous but non-smooth behavior, and multivalued behavior, respectively. This indicates that these entanglement properties have the ability to characterize phase transitions in the standard coordinates. Additionally, we notice that insignificant changes occur in these entanglement properties in the low-temperature phase. However, as the temperature approaches the phase boundary, these properties undergo significant decreases. This difference lies in the fact that the conditional mutual information ($CMI$) exhibits a monotonic decrease in the high-temperature region, while both $S_{EE}^{ren}$ and $EoP$ change from decreasing to increasing with rising temperature, leading to the presence of a minimum value in the high-temperature region. Furthermore, these entanglement quantities exhibit consistent behavior across the two EMD models.

In fact, it has been found that QCD matter is strongly coupled at temperatures below $300$ MeV~\cite{Shuryak:2014zxa}. While, when temperatures surpass $300$ MeV, the results from hard thermal loop perturbation theory in QCD and lattice QCD simulations align~\cite{Ghiglieri:2020dpq}. This suggests that above $300$ MeV, QCD matter may transition to a weakly coupled state, making it unsuitable for studying the properties of its dual QCD matter using weakly coupled gravity. This poses a challenge in applying the gauge/gravity duality, known for its strong-weak coupling correspondence, to explore the properties of QCD matter at high temperature. Given that the purpose of this work is to investigate the behavior of entanglement on QCD phase diagrams, if one only focus on the behavior of entanglement near the phase boundary ($T<200~\text{MeV}$), it is clearly that in standard coordinates, the entanglement entropy, conditional mutual information, and entanglement of purification exhibit consistent patterns. This observation indicates that these three types of cutoff-independent entanglement measures offer a robust description of the entanglement between different subregions in boundary field theory: entanglement properties decrease as the temperature increases. Specifically, entanglement properties remain relatively constant in the low-temperature phase, start to diminish near the phase boundary, and further decrease in the high-temperature phase until the temperature approaches $300$ MeV. 

\section{Conclusion}\label{sec:05}
The objective of this study is to identify potential signatures of entanglement properties near phase boundary in holographic QCD models. Two holographic EMD models were considered, and various entanglement measures, such as entanglement entropy, conditional mutual information, and entanglement of purification, were investigated. The behaviors of these entanglement properties were analyzed and compared between the two models. We observed that the entanglement entropy $S_{EE}^{reg}$, obtained through the direct UV-cutoff renormalization method, showed significant dependence on the specific model employed. Other non-divergent entanglement measures, such as cutoff-independent entanglement entropy $S_{EE}^{ren}$, conditional mutual information $CMI$, and entanglement of purification $EoP$, exhibited similar behaviors in different models.

Furthermore, we examined the behavior of $S_{EE}^{ren}$, $CMI$, and $EoP$ in standard coordinates. We found that the behavior of different entanglement properties varies at high temperature region ($300 ~\text{MeV}\leq T$). The cutoff-independent entanglement entropy and the entanglement of purification display non-monotonic behavior at high temperatures. However, the conditional mutual information consistently exhibits a monotonically decreasing trend. It is important to note that due to the possibility of weak coupling state of QCD matter at high temperature, the applicability of holographic methods for describing the properties of high-temperature QCD matter may be limited. While, when we focus specifically on the temperature range below $300$ MeV, we observe a similar behavior among the three entanglement measures. This suggests that within this temperature range, the entanglement of QCD matter decreases with increasing temperature. We also found that entanglement properties exhibited distinct phase transition behavior at the critical end point (CEP) and first-order phase transitions. 

Note that while entanglement properties can effectively serve as a phase transition order parameter and exhibit distinguishable signals at critical endpoint (CEP) and first-order phase transitions, it may not exhibit clear phase transition signals in the crossover region. Analyzing the characteristic signals of crossover phase transitions based on the behavior of entanglement properties is an area of focus for our future research.

Another noteworthy concern is that The EMD model yields that the cutoff-independent entanglement properties exhibit trivial behavior at low temperatures, with their values remaining nearly constant. This could be attributed to the EMD model's inability to capture the physics of the low-temperature phase dominated by hadronic matter. Furthermore, the soft-wall model typically necessitates a quadratic dilaton field to align with the characteristics of hadronic and gluonic matter, a criterion that the dilaton field derived from the EMD model often fails to satisfy. This discrepancy implies that while the EMD model provides a plausible depiction of the QCD equation of state near phase boundaries $(100 ~\text{MeV}<T<300~\text{MeV})$, its applicability at lower (or higher) temperatures is questionable. In the forthcoming model development, it is imperative to leverage the distinctive features of both models to achieve an accurate holographic depiction of QCD matter over a wider range of temperatures and densities.


\section{Acknowledgments}\label{sec:06}

We would like to thank Yong Cai, Song He, Li Li and Peng Liu, for their helpful discussion. We acknowledge support from the national Key Program for Science and Technology Research Development (2023YFB3002500).

\begin{appendix}
\section{Entanglement properties in numerical coordinates}\label{sec:041}
In principle, all physical quantities, including entanglement properties, should be computed in standard coordinates. Nevertheless, the behavior of entanglement properties in numerical coordinates has also received attention \cite{Knaute:2017lll}. Exploring the similarities and differences in the behavior of entanglement properties between numerical coordinates and standard coordinates holds theoretical significance. Therefore, in this section, we demonstrate the behavior of entanglement properties between infinitely long strip regions on the boundary of a spacetime background using the metric of numerical coordinates.

The holographic entanglement entropy of Model I in numerical coordinates can be calculated as
\begin{equation}\label{eqa1}
  S_{EE}=\frac{2\pi}{\kappa_N^2}\int dx_1 dx_2 dx_3 \sqrt{\gamma}
=\frac{2\pi V_2}{\kappa_N^2}\int_{-\ell/2}^{\ell/2} dx_1 r^2 \sqrt{r^2+\frac{r'(x_1)^2}{f(r)}},
\end{equation}
and the conserved quantity
\begin{equation}\label{eqa2}
  \frac{r^4}{\sqrt{r^2+\frac{r'(x_1)^2}{f(r)}}}=r_{*}^3 ~~~ \Rightarrow ~~~ r'(x_1)=\frac{r}{r_*^3}\sqrt{f(r)(r^6-r_*^6)}
\end{equation}
with $r_*$ the minimum value of $r$ on the R-T surface. The length of the interval shows
\begin{equation}\label{eqa3}
\ell=2\int_{r_*}^{\infty}r'(x_1)^{-1}dr=2\int_{r_*}^{\infty}\frac{r_*^3}{r\sqrt{f(r)(r^6-r_*^6)}}dr.
\end{equation}
Then, the expressions for the holographic entanglement entropy and the entanglement of purification are given as follows
\begin{equation}\label{eqa4}
    S_{EE}=\frac{4\pi V_2}{\kappa_N^2}\int_{r_*}^{\infty}\frac{r^5}{\sqrt{f(r)(r^6-r_*^6)}}dr,
\end{equation}
and
\begin{equation}\label{eqa5}
  EOP=\frac{2\pi}{\kappa_N^2}\int dr dx_2 dx_3 \sqrt{\sigma}=\frac{2\pi V_2}{\kappa_N^2}\int_{r_{1*}}^{r_{2*}} r^2 \frac{1}{\sqrt{f(r)}} dr.
\end{equation}
Here, $r_{1*}$ and $r_{2*}$ refer to the $r_*$ values associated with the R-T surfaces of $B$ and $ABC$, respectively. 

The holographic entanglement entropy for Model II in numerical coordinates can be calculated using the following formula
\begin{equation}\label{eqa6}
  S_{EE}=\frac{4\pi V_2 }{\kappa_N^2}\int_{-\ell/2}^{\ell/2} dx_1 e^{2 A(r)} \sqrt{e^{2 A(r)}+\frac{r'(x_1)^2}{h(r)}},
\end{equation}
and the conserved quantity
\begin{equation}\label{eqa7}
  \frac{e^{4 A(r)}}{\sqrt{e^{2 A(r)}+\frac{ r'(x_1)^2}{h(r)}}}=e^{3 A(r_*)} ~~~\Rightarrow~~~ r'(x_1)=e^{A(r)-3A(r_*)}\sqrt{(e^{6A(r)}-e^{6A(r_*)})h(r)}.
\end{equation}
Then the length of the interval is
\begin{equation}\label{eqa8}
\ell=2\int_{r_*}^{\infty}r'(x_1)^{-1}dr=2\int_{r_*}^{\infty}\frac{e^{3A(r_*)-A(r)}}{\sqrt{(e^{6A(r)}-e^{6A(r_*)})h(r)}}dr.
\end{equation}
And the holographic entanglement entropy and the entanglement of purification in Model II are as
\begin{equation}\label{eqa9}
  S_{EE}=\frac{4\pi V_2 }{\kappa_N^2}\int_{r_*}^{\infty}\frac{ e^{5A(r)}}{\sqrt{(e^{6A(r)}-e^{6A(r_*)})h(r)}}dr,
\end{equation}
and
\begin{equation}\label{eqa10}
  EOP=\frac{2\pi}{\kappa_N^2}\int dr dx_2 dx_3 \sqrt{\sigma}=\frac{2\pi V_2 }{\kappa_N^2}\int_{r_{1*}}^{r_{2*}}  \frac{ e^{2A(r)}}{\sqrt{h(r)}} dr.
\end{equation}

The renormalized entanglement entropies are denoted as $S_{EE}^{reg}$ and $S_{EE}^{ren}$, respectively, and employ the same renormalization scheme as in standard coordinates. In Model I, we set the UV-cutoff value to $r=\tilde{r}=200$. In Model II, we use UV-cutoff values of $r=2$ in numerical coordinates. Similar to calculations in standard coordinates, we fix the 2 dimensional volume term as $V=\frac{\kappa_N^2}{2\pi}$ in numerical coordinates.

\begin{figure}[htbp]
\centering
\includegraphics[width=.485\textwidth]{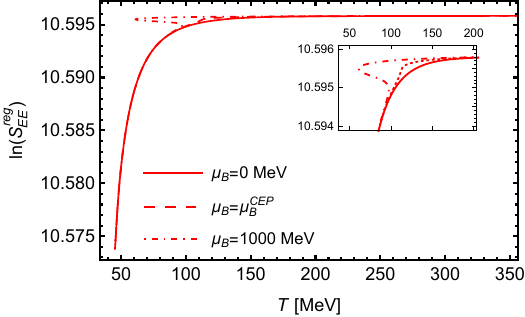}
\qquad
\includegraphics[width=.455\textwidth]{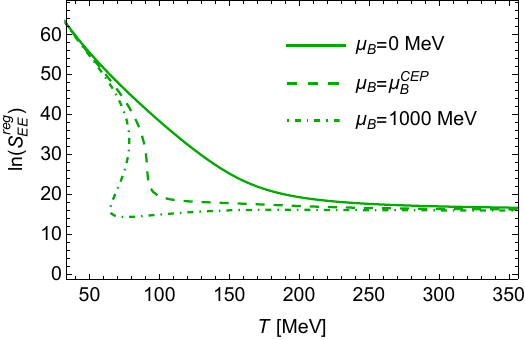}
  \caption{The behaviors of entanglement entropy $\ln(S_{EE}^{reg})$ at $\mu_B=0,~\mu_B^{CEP}~\text{and}~1000~\text{MeV}$ from Model I (\textbf{Left panel}) and Model II (\textbf{Right panel}) in numerical coordinates. Note that the behavior of $S_{EE}^{reg}$ from Model II is consistent with that in \cite{Knaute:2017lll}, which used a similar model and both in numerical coordinates.}\label{seeo}
\end{figure}

Fig.~\ref{seeo} plots the behaviors of the entanglement entropy $\ln(S_{EE}^{reg})$ in numerical coordinates for Model I and Model II at $\mu_B=0,~\mu_B^{CEP},$ and $1000~\text{MeV}$. In Model I, at $\mu_B=0$, the entanglement entropy $\ln(S_{EE}^{reg})$ shows a monotonically increasing trend with increasing temperature. At $\mu_B=\mu_B^{CEP}$, as the temperature rises, $\ln(S_{EE}^{reg})$ exhibits a steady increase, but there is a sharp rise at the phase transition temperature. For $\mu_B>\mu_B^{CEP}$, the entanglement entropy displays a multivalued behavior near the phase transition temperature. Similar monotonic, rapid increasing, and multivalued behaviors are also observed in Model II, as shown in the right panel of Fig.~\ref{seeo}. Furthermore, in the low-temperature phase, the change in $\ln(S_{EE}^{reg})$ is more pronounced, while at high temperatures, the $\ln(S_{EE}^{reg})$ values for different chemical potentials tend to converge. However, it is crucial to note that the increase and decrease patterns of $\ln(S_{EE}^{reg})$ exhibit significant dependence on the specific metric used. In particular, $\ln(S_{EE}^{reg})$ displays a decreasing trend with increasing temperature in Model II, highlighting the influence of the metric's specific form on the behavior of the entanglement entropy. It is worth mentioning that the behavior of $S_{EE}^{reg}$ from Model II is consistent with that in \cite{Knaute:2017lll}, which used a similar model and both in numerical coordinates.

\begin{figure}[htbp]
\centering
\includegraphics[width=.98\textwidth]{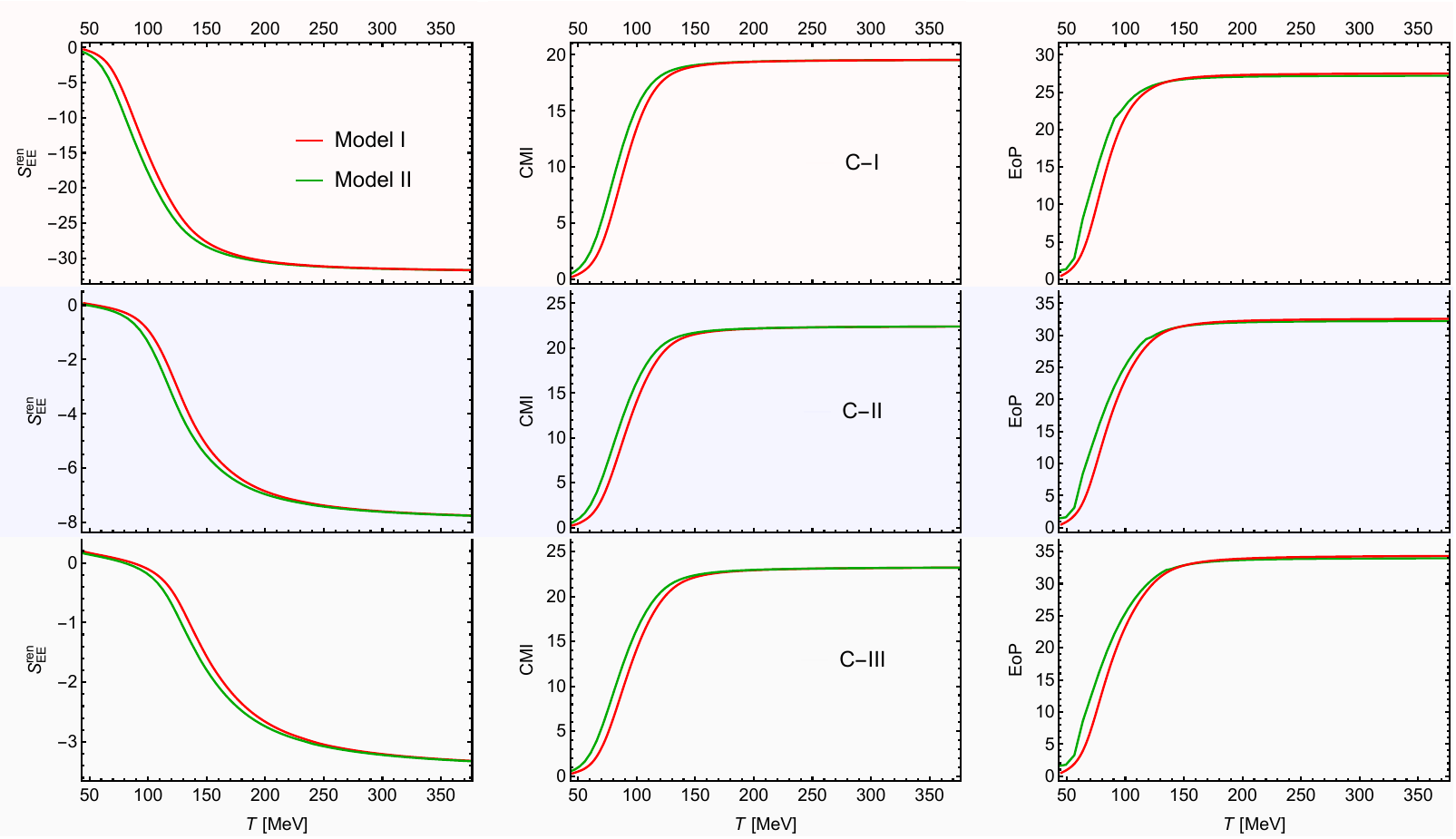}
  \caption{The behaviors of entanglement entropy $S_{EE}^{ren}$, conditional mutual information $CMI$ and entanglement of purification $EoP$ for different configurations of intervals in numerical coordinates at $\mu_B=300~\text{MeV}$. The top, middle, and bottom rows of figures correspond to three different interval configurations: C-I, C-II, and C-III, respectively, as presented in Table~\ref{table3}.}\label{orifig0}
\end{figure}

To explore the model dependence of non-divergent entanglement properties, we examine the behavior of $S_{EE}^{ren}$, $CMI$, and $EoP$ with temperature at $\mu_B=300~\text{MeV}$ using various interval configurations (as depicted in Table~\ref{table3}), as shown in Fig.\ref{orifig0}.
\begin{table}[htbp]
\centering
\begin{tabular}{|c|c|c|c|}
\hline
                 &  C-I  &  C-II  & C-III   \\ \hline
    $S_{EE}^{ren}$  &  $\ell=0.1$  &  $\ell=0.2$  & $\ell=0.3$    \\ \hline
    $CMI$          &  $\ell=0.1, ~0.1, ~0.1$  &  $\ell=0.1, ~0.1, ~0.2$  &  $\ell=0.1, ~0.1, ~0.3$   \\ \hline
    $EoP$    &   $\ell=0.1, ~0.2$  & $\ell=0.2, ~0.3$   &  $\ell=0.3, ~0.4$      \\ \hline
\end{tabular}
\caption{The three different configurations of intervals in numerical coordinates for Fig.~\ref{orifig0}.}
    \label{table3}
\end{table}

Fig.~\ref{orifig0} compare the entanglement entropy $S_{EE}^{ren}$, conditional mutual information $CMI$, and entanglement of purification $EoP$ at $\mu_B=300~\text{MeV}$ for different interval configurations. The behavior of these entanglement properties remains largely consistent between Model I and Model II, with minor numerical discrepancies. Furthermore, the qualitative behavior of these entanglement quantities remains relatively unchanged across different interval settings, with only quantitative variations. Moreover, Fig.~\ref{orifig0} demonstrates that the entanglement properties show a monotonic behavior with temperature for $\mu_B=300~\text{MeV}$. Comparing different rows, we observe that the entanglement properties display similar patterns for different interval configurations, differing primarily in numerical values.

\begin{figure}[htbp]
\centering
\includegraphics[width=.98\textwidth]{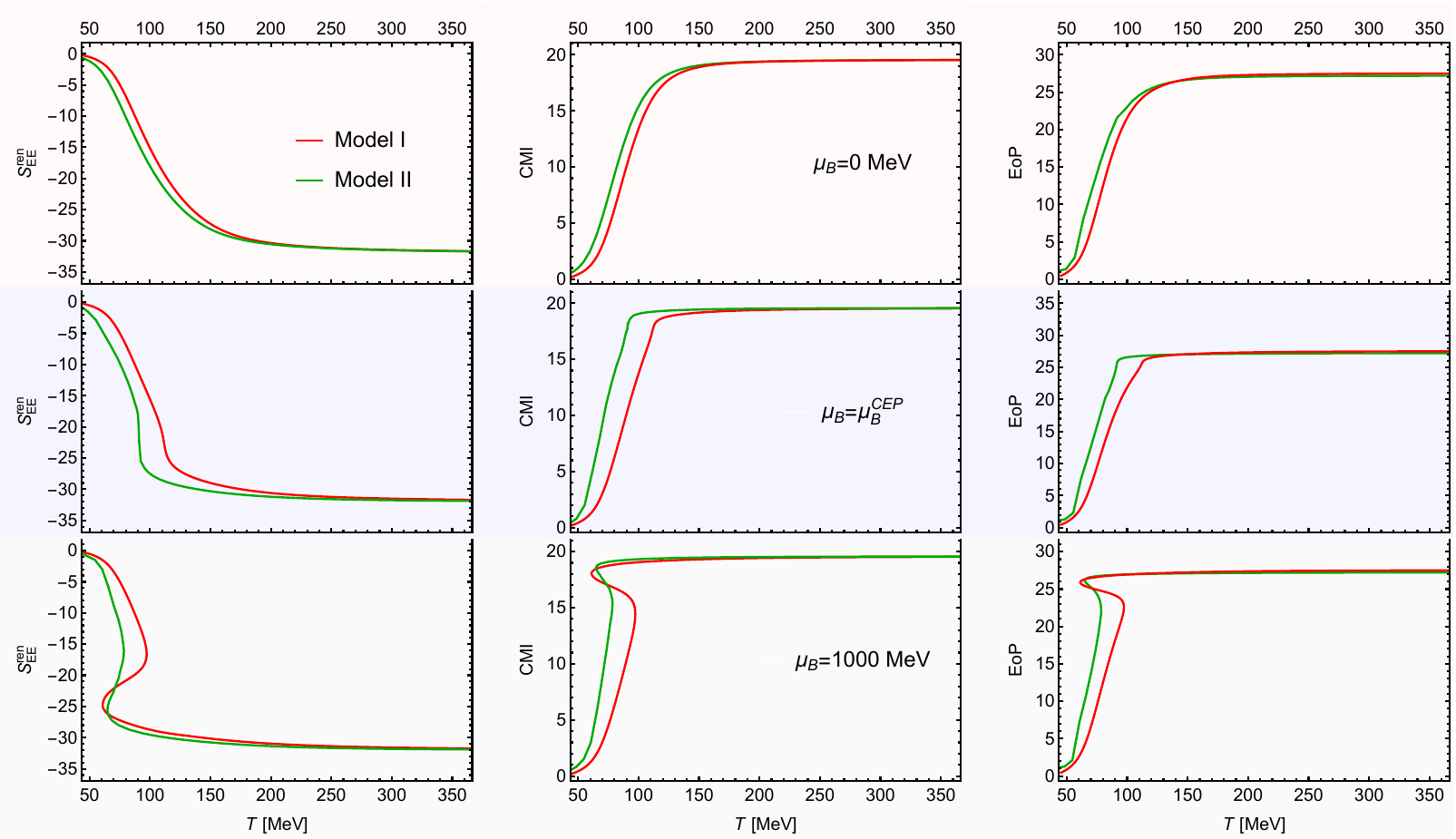}
  \caption{The universal behaviors of entanglement properties $S_{EE}^{ren}$, $CMI$ and $EoP$ in numerical coordinates at different chemical potential $\mu_B$. The top, middle and bottom three figures are at $\mu_B=0,~\mu_B^{CEP}~\text{and}~1000~\text{MeV}$, respectively.}\label{orifig}
\end{figure}

Fig.~\ref{orifig} shows the behavior of different entanglement properties in the crossover region ($\mu_B=0$), the critical endpoint (CEP) ($\mu_B=\mu_B^{CEP}$), and the first-order phase transition region ($\mu_B=1000~\text{MeV}$). The figure reveals three noteworthy features: 1) In the low temperature phase, these entanglement properties approach zero, while in the high temperature phase, they tend to converge to a fixed value. 2) There are distinct characteristic behaviors at the phase boundaries. In the crossover region, these entanglement properties exhibit single-valued and smooth behavior, with a significant increase near the phase boundary. At CEP, these properties are also monotonic but not smooth at the phase boundary. In the first-order phase transition region, the entanglements display multi-valued behavior. 3) The results obtained from both models exhibit nearly identical behavior, with only minor quantitative differences. 
\end{appendix}


\begin{thebibliography}{100}
	
	\bibitem{Borsanyi:2010cj}
	S.~Borsanyi, G.~Endrodi, Z.~Fodor, A.~Jakovac, S.~D. Katz, S.~Krieg, C.~Ratti,
	and K.~K. Szabo, ``{The QCD equation of state with dynamical quarks},''
	\href{http://dx.doi.org/10.1007/JHEP11(2010)077}{{\em JHEP} {\bfseries 11}
		(2010) 077}, \href{http://arxiv.org/abs/1007.2580}{{\ttfamily arXiv:1007.2580
			[hep-lat]}}.
	
	\bibitem{Borsanyi:2013bia}
	S.~Borsanyi, Z.~Fodor, C.~Hoelbling, S.~D. Katz, S.~Krieg, and K.~K. Szabo,
	``{Full result for the QCD equation of state with 2+1 flavors},''
	\href{http://dx.doi.org/10.1016/j.physletb.2014.01.007}{{\em Phys. Lett. B}
		{\bfseries 730} (2014) 99--104},
	\href{http://arxiv.org/abs/1309.5258}{{\ttfamily arXiv:1309.5258 [hep-lat]}}.
	
	\bibitem{HotQCD:2014kol}
	{\bfseries HotQCD} Collaboration, A.~Bazavov {\em et~al.}, ``{Equation of state
		in ( 2+1 )-flavor QCD},''
	\href{http://dx.doi.org/10.1103/PhysRevD.90.094503}{{\em Phys. Rev. D}
		{\bfseries 90} (2014) 094503},
	\href{http://arxiv.org/abs/1407.6387}{{\ttfamily arXiv:1407.6387 [hep-lat]}}.
	
	\bibitem{Xin:2014ela}
	X.-y. Xin, S.-x. Qin, and Y.-x. Liu, ``{Quark number fluctuations at finite
		temperature and finite chemical potential via the Dyson-Schwinger equation
		approach},'' \href{http://dx.doi.org/10.1103/PhysRevD.90.076006}{{\em Phys.
			Rev. D} {\bfseries 90} no.~7, (2014) 076006},
	\href{http://arxiv.org/abs/2109.09935}{{\ttfamily arXiv:2109.09935
			[hep-ph]}}.
	
	\bibitem{Gao:2016qkh}
	F.~Gao and Y.-x. Liu, ``{QCD phase transitions via a refined truncation of
		Dyson-Schwinger equations},''
	\href{http://dx.doi.org/10.1103/PhysRevD.94.076009}{{\em Phys. Rev. D}
		{\bfseries 94} no.~7, (2016) 076009},
	\href{http://arxiv.org/abs/1607.01675}{{\ttfamily arXiv:1607.01675
			[hep-ph]}}.
	
	\bibitem{Qin:2010nq}
	S.-x. Qin, L.~Chang, H.~Chen, Y.-x. Liu, and C.~D. Roberts, ``{Phase diagram
		and critical endpoint for strongly-interacting quarks},''
	\href{http://dx.doi.org/10.1103/PhysRevLett.106.172301}{{\em Phys. Rev.
			Lett.} {\bfseries 106} (2011) 172301},
	\href{http://arxiv.org/abs/1011.2876}{{\ttfamily arXiv:1011.2876 [nucl-th]}}.
	
	\bibitem{Shi:2014zpa}
	C.~Shi, Y.-L. Wang, Y.~Jiang, Z.-F. Cui, and H.-S. Zong, ``{Locate QCD Critical
		End Point in a Continuum Model Study},''
	\href{http://dx.doi.org/10.1007/JHEP07(2014)014}{{\em JHEP} {\bfseries 07}
		(2014) 014}, \href{http://arxiv.org/abs/1403.3797}{{\ttfamily arXiv:1403.3797
			[hep-ph]}}.
	
	\bibitem{Fischer:2014ata}
	C.~S. Fischer, J.~Luecker, and C.~A. Welzbacher, ``{Phase structure of three
		and four flavor QCD},''
	\href{http://dx.doi.org/10.1103/PhysRevD.90.034022}{{\em Phys. Rev. D}
		{\bfseries 90} no.~3, (2014) 034022},
	\href{http://arxiv.org/abs/1405.4762}{{\ttfamily arXiv:1405.4762 [hep-ph]}}.
	
	\bibitem{Gao:2020qsj}
	F.~Gao and J.~M. Pawlowski, ``{QCD phase structure from functional methods},''
	\href{http://dx.doi.org/10.1103/PhysRevD.102.034027}{{\em Phys. Rev. D}
		{\bfseries 102} no.~3, (2020) 034027},
	\href{http://arxiv.org/abs/2002.07500}{{\ttfamily arXiv:2002.07500
			[hep-ph]}}.
	
	\bibitem{Asakawa:1989bq}
	M.~Asakawa and K.~Yazaki, ``{Chiral Restoration at Finite Density and
		Temperature},'' \href{http://dx.doi.org/10.1016/0375-9474(89)90002-X}{{\em
			Nucl. Phys. A} {\bfseries 504} (1989) 668--684}.
	
	\bibitem{Schwarz:1999dj}
	T.~M. Schwarz, S.~P. Klevansky, and G.~Papp, ``{The Phase diagram and bulk
		thermodynamical quantities in the NJL model at finite temperature and
		density},'' \href{http://dx.doi.org/10.1103/PhysRevC.60.055205}{{\em Phys.
			Rev. C} {\bfseries 60} (1999) 055205},
	\href{http://arxiv.org/abs/nucl-th/9903048}{{\ttfamily
			arXiv:nucl-th/9903048}}.
	
	\bibitem{Li:2018ygx}
	Z.~Li, K.~Xu, X.~Wang, and M.~Huang, ``{The kurtosis of net baryon number
		fluctuations from a realistic
		Polyakov\textendash{}Nambu\textendash{}Jona-Lasinio model along the
		experimental freeze-out line},''
	\href{http://dx.doi.org/10.1140/epjc/s10052-019-6703-x}{{\em Eur. Phys. J. C}
		{\bfseries 79} no.~3, (2019) 245},
	\href{http://arxiv.org/abs/1801.09215}{{\ttfamily arXiv:1801.09215
			[hep-ph]}}.
	
	\bibitem{Zhuang:2000ub}
	P.~Zhuang, M.~Huang, and Z.~Yang, ``{Density effect on hadronization of a quark
		plasma},'' \href{http://dx.doi.org/10.1103/PhysRevC.62.054901}{{\em Phys.
			Rev. C} {\bfseries 62} (2000) 054901},
	\href{http://arxiv.org/abs/nucl-th/0008043}{{\ttfamily
			arXiv:nucl-th/0008043}}.
	
	\bibitem{Fu:2019hdw}
	W.-j. Fu, J.~M. Pawlowski, and F.~Rennecke, ``{QCD phase structure at finite
		temperature and density},''
	\href{http://dx.doi.org/10.1103/PhysRevD.101.054032}{{\em Phys. Rev. D}
		{\bfseries 101} no.~5, (2020) 054032},
	\href{http://arxiv.org/abs/1909.02991}{{\ttfamily arXiv:1909.02991
			[hep-ph]}}.
	
	\bibitem{Zhang:2017icm}
	H.~Zhang, D.~Hou, T.~Kojo, and B.~Qin, ``{Functional renormalization group
		study of the quark-meson model with $\omega$ meson},''
	\href{http://dx.doi.org/10.1103/PhysRevD.96.114029}{{\em Phys. Rev. D}
		{\bfseries 96} no.~11, (2017) 114029},
	\href{http://arxiv.org/abs/1709.05654}{{\ttfamily arXiv:1709.05654
			[hep-ph]}}.
	
	\bibitem{Fu:2021oaw}
	W.-j. Fu, X.~Luo, J.~M. Pawlowski, F.~Rennecke, R.~Wen, and S.~Yin,
	``{Hyper-order baryon number fluctuations at finite temperature and
		density},'' \href{http://dx.doi.org/10.1103/PhysRevD.104.094047}{{\em Phys.
			Rev. D} {\bfseries 104} no.~9, (2021) 094047},
	\href{http://arxiv.org/abs/2101.06035}{{\ttfamily arXiv:2101.06035
			[hep-ph]}}.
	
	\bibitem{Vovchenko:2017gkg}
	V.~Vovchenko, J.~Steinheimer, O.~Philipsen, and H.~Stoecker, ``{Cluster
		Expansion Model for QCD Baryon Number Fluctuations: No Phase Transition at
		$\mu_B / T < \pi$},''
	\href{http://dx.doi.org/10.1103/PhysRevD.97.114030}{{\em Phys. Rev. D}
		{\bfseries 97} no.~11, (2018) 114030},
	\href{http://arxiv.org/abs/1711.01261}{{\ttfamily arXiv:1711.01261
			[hep-ph]}}.
	
	\bibitem{Borsanyi:2020fev}
	S.~Borsanyi, Z.~Fodor, J.~N. Guenther, R.~Kara, S.~D. Katz, P.~Parotto,
	A.~Pasztor, C.~Ratti, and K.~K. Szabo, ``{QCD Crossover at Finite Chemical
		Potential from Lattice Simulations},''
	\href{http://dx.doi.org/10.1103/PhysRevLett.125.052001}{{\em Phys. Rev.
			Lett.} {\bfseries 125} no.~5, (2020) 052001},
	\href{http://arxiv.org/abs/2002.02821}{{\ttfamily arXiv:2002.02821
			[hep-lat]}}.
	
	\bibitem{Bazavov:2020bjn}
	A.~Bazavov {\em et~al.}, ``{Skewness, kurtosis, and the fifth and sixth order
		cumulants of net baryon-number distributions from lattice QCD confront
		high-statistics STAR data},''
	\href{http://dx.doi.org/10.1103/PhysRevD.101.074502}{{\em Phys. Rev. D}
		{\bfseries 101} no.~7, (2020) 074502},
	\href{http://arxiv.org/abs/2001.08530}{{\ttfamily arXiv:2001.08530
			[hep-lat]}}.
	
	\bibitem{Borsanyi:2021sxv}
	S.~Bors\'anyi, Z.~Fodor, J.~N. Guenther, R.~Kara, S.~D. Katz, P.~Parotto,
	A.~P\'asztor, C.~Ratti, and K.~K. Szab\'o, ``{Lattice QCD equation of state
		at finite chemical potential from an alternative expansion scheme},''
	\href{http://dx.doi.org/10.1103/PhysRevLett.126.232001}{{\em Phys. Rev.
			Lett.} {\bfseries 126} no.~23, (2021) 232001},
	\href{http://arxiv.org/abs/2102.06660}{{\ttfamily arXiv:2102.06660
			[hep-lat]}}.
	
	\bibitem{Bollweg:2022fqq}
	D.~Bollweg, D.~A. Clarke, J.~Goswami, O.~Kaczmarek, F.~Karsch, S.~Mukherjee,
	P.~Petreczky, C.~Schmidt, and S.~Sharma, ``{Equation of state and speed of
		sound of (2+1)-flavor QCD in strangeness-neutral matter at non-vanishing net
		baryon-number density},'' \href{http://arxiv.org/abs/2212.09043}{{\ttfamily
			arXiv:2212.09043 [hep-lat]}}.
	
	\bibitem{Philipsen:2021qji}
	O.~Philipsen, ``{Lattice Constraints on the QCD Chiral Phase Transition at
		Finite Temperature and Baryon Density},''
	\href{http://dx.doi.org/10.3390/sym13112079}{{\em Symmetry} {\bfseries 13}
		no.~11, (2021) 2079}, \href{http://arxiv.org/abs/2111.03590}{{\ttfamily
			arXiv:2111.03590 [hep-lat]}}.
	
	\bibitem{Vidal:2002zz}
	G.~Vidal and R.~F. Werner, ``{Computable measure of entanglement},''
	\href{http://dx.doi.org/10.1103/PhysRevA.65.032314}{{\em Phys. Rev. A}
		{\bfseries 65} (2002) 032314},
	\href{http://arxiv.org/abs/quant-ph/0102117}{{\ttfamily
			arXiv:quant-ph/0102117}}.
	
	\bibitem{Plenio:2005cwa}
	M.~B. Plenio, ``{Logarithmic Negativity: A Full Entanglement Monotone That is
		not Convex},'' \href{http://dx.doi.org/10.1103/PhysRevLett.95.090503}{{\em
			Phys. Rev. Lett.} {\bfseries 95} (2005) 090503},
	\href{http://arxiv.org/abs/quant-ph/0505071}{{\ttfamily
			arXiv:quant-ph/0505071}}.
	
	\bibitem{Horodecki:2009zz}
	R.~Horodecki, P.~Horodecki, M.~Horodecki, and K.~Horodecki, ``{Quantum
		entanglement},'' \href{http://dx.doi.org/10.1103/RevModPhys.81.865}{{\em Rev.
			Mod. Phys.} {\bfseries 81} (2009) 865--942},
	\href{http://arxiv.org/abs/quant-ph/0702225}{{\ttfamily
			arXiv:quant-ph/0702225}}.
	
	\bibitem{Jokela:2020wgs}
	N.~Jokela and J.~G. Subils, ``{Is entanglement a probe of confinement?},''
	\href{http://dx.doi.org/10.1007/JHEP02(2021)147}{{\em JHEP} {\bfseries 02}
		(2021) 147}, \href{http://arxiv.org/abs/2010.09392}{{\ttfamily
			arXiv:2010.09392 [hep-th]}}.
	
	\bibitem{Ryu:2006bv}
	S.~Ryu and T.~Takayanagi, ``{Holographic derivation of entanglement entropy
		from AdS/CFT},'' \href{http://dx.doi.org/10.1103/PhysRevLett.96.181602}{{\em
			Phys. Rev. Lett.} {\bfseries 96} (2006) 181602},
	\href{http://arxiv.org/abs/hep-th/0603001}{{\ttfamily arXiv:hep-th/0603001}}.
	
	\bibitem{Ryu:2006ef}
	S.~Ryu and T.~Takayanagi, ``{Aspects of Holographic Entanglement Entropy},''
	\href{http://dx.doi.org/10.1088/1126-6708/2006/08/045}{{\em JHEP} {\bfseries
			08} (2006) 045}, \href{http://arxiv.org/abs/hep-th/0605073}{{\ttfamily
			arXiv:hep-th/0605073}}.
	
	\bibitem{Nishioka:2009un}
	T.~Nishioka, S.~Ryu, and T.~Takayanagi, ``{Holographic Entanglement Entropy: An
		Overview},'' \href{http://dx.doi.org/10.1088/1751-8113/42/50/504008}{{\em J.
			Phys. A} {\bfseries 42} (2009) 504008},
	\href{http://arxiv.org/abs/0905.0932}{{\ttfamily arXiv:0905.0932 [hep-th]}}.
	
	\bibitem{Rangamani:2016dms}
	M.~Rangamani and T.~Takayanagi,
	\href{http://dx.doi.org/10.1007/978-3-319-52573-0}{{\em {Holographic
				Entanglement Entropy}}}, vol.~931.
	\newblock Springer, 2017.
	\newblock \href{http://arxiv.org/abs/1609.01287}{{\ttfamily arXiv:1609.01287
			[hep-th]}}.
	
	\bibitem{Takayanagi:2017knl}
	T.~Takayanagi and K.~Umemoto, ``{Entanglement of purification through
		holographic duality},''
	\href{http://dx.doi.org/10.1038/s41567-018-0075-2}{{\em Nature Phys.}
		{\bfseries 14} no.~6, (2018) 573--577},
	\href{http://arxiv.org/abs/1708.09393}{{\ttfamily arXiv:1708.09393
			[hep-th]}}.
	
	\bibitem{Zhang:2016rcm}
	S.-J. Zhang, ``{Holographic entanglement entropy close to crossover/phase
		transition in strongly coupled systems},''
	\href{http://dx.doi.org/10.1016/j.nuclphysb.2017.01.010}{{\em Nucl. Phys. B}
		{\bfseries 916} (2017) 304--319},
	\href{http://arxiv.org/abs/1608.03072}{{\ttfamily arXiv:1608.03072
			[hep-th]}}.
	
	\bibitem{Ali-Akbari:2017vtb}
	M.~Ali-Akbari and M.~Lezgi, ``{Holographic QCD, entanglement entropy, and
		critical temperature},''
	\href{http://dx.doi.org/10.1103/PhysRevD.96.086014}{{\em Phys. Rev. D}
		{\bfseries 96} no.~8, (2017) 086014},
	\href{http://arxiv.org/abs/1706.04335}{{\ttfamily arXiv:1706.04335
			[hep-th]}}.
	
	\bibitem{Knaute:2017lll}
	J.~Knaute and B.~K\"ampfer, ``{Holographic Entanglement Entropy in the QCD
		Phase Diagram with a Critical Point},''
	\href{http://dx.doi.org/10.1103/PhysRevD.96.106003}{{\em Phys. Rev. D}
		{\bfseries 96} no.~10, (2017) 106003},
	\href{http://arxiv.org/abs/1706.02647}{{\ttfamily arXiv:1706.02647
			[hep-ph]}}.
	
	\bibitem{Dudal:2018ztm}
	D.~Dudal and S.~Mahapatra, ``{Interplay between the holographic QCD phase
		diagram and entanglement entropy},''
	\href{http://dx.doi.org/10.1007/JHEP07(2018)120}{{\em JHEP} {\bfseries 07}
		(2018) 120}, \href{http://arxiv.org/abs/1805.02938}{{\ttfamily
			arXiv:1805.02938 [hep-th]}}.
	
	\bibitem{Li:2020pgn}
	Z.~Li, K.~Xu, and M.~Huang, ``{The entanglement properties of holographic QCD
		model with a critical end point},''
	\href{http://dx.doi.org/10.1088/1674-1137/abc539}{{\em Chin. Phys. C}
		{\bfseries 45} no.~1, (2021) 013116},
	\href{http://arxiv.org/abs/2002.08650}{{\ttfamily arXiv:2002.08650
			[hep-th]}}.
	
	\bibitem{Asadi:2022mvo}
	M.~Asadi, B.~Amrahi, and H.~Eshaghi-Kenari, ``{Probing phase structure of
		strongly coupled matter with holographic entanglement measures},''
	\href{http://dx.doi.org/10.1140/epjc/s10052-023-11214-6}{{\em Eur. Phys. J.
			C} {\bfseries 83} no.~1, (2023) 69},
	\href{http://arxiv.org/abs/2209.01586}{{\ttfamily arXiv:2209.01586
			[hep-th]}}.
	
	\bibitem{Yadav:2022mnv}
	G.~Yadav and A.~Misra, ``{Entanglement entropy and Page curve from the M-theory
		dual of thermal QCD above Tc at intermediate coupling},''
	\href{http://dx.doi.org/10.1103/PhysRevD.107.106015}{{\em Phys. Rev. D}
		{\bfseries 107} no.~10, (2023) 106015},
	\href{http://arxiv.org/abs/2207.04048}{{\ttfamily arXiv:2207.04048
			[hep-th]}}.
	
	\bibitem{Takayanagi:2014rue}
	T.~Takayanagi, ``{Entanglement entropy and gravity/condensed matter
		correspondence},'' \href{http://dx.doi.org/10.1007/s10714-014-1693-3}{{\em
			Gen. Rel. Grav.} {\bfseries 46} (2014) 1693}.
	
	\bibitem{Ling:2015dma}
	Y.~Ling, P.~Liu, C.~Niu, J.-P. Wu, and Z.-Y. Xian, ``{Holographic Entanglement
		Entropy Close to Quantum Phase Transitions},''
	\href{http://dx.doi.org/10.1007/JHEP04(2016)114}{{\em JHEP} {\bfseries 04}
		(2016) 114}, \href{http://arxiv.org/abs/1502.03661}{{\ttfamily
			arXiv:1502.03661 [hep-th]}}.
	
	\bibitem{Ling:2016dck}
	Y.~Ling, P.~Liu, J.-P. Wu, and Z.~Zhou, ``{Holographic Metal-Insulator
		Transition in Higher Derivative Gravity},''
	\href{http://dx.doi.org/10.1016/j.physletb.2016.12.051}{{\em Phys. Lett. B}
		{\bfseries 766} (2017) 41--48},
	\href{http://arxiv.org/abs/1606.07866}{{\ttfamily arXiv:1606.07866
			[hep-th]}}.
	
	\bibitem{Ling:2016wyr}
	Y.~Ling, P.~Liu, and J.-P. Wu, ``{Characterization of Quantum Phase Transition
		using Holographic Entanglement Entropy},''
	\href{http://dx.doi.org/10.1103/PhysRevD.93.126004}{{\em Phys. Rev. D}
		{\bfseries 93} no.~12, (2016) 126004},
	\href{http://arxiv.org/abs/1604.04857}{{\ttfamily arXiv:1604.04857
			[hep-th]}}.
	
	\bibitem{Cai:2012sk}
	R.-G. Cai, S.~He, L.~Li, and Y.-L. Zhang, ``{Holographic Entanglement Entropy
		in Insulator/Superconductor Transition},''
	\href{http://dx.doi.org/10.1007/JHEP07(2012)088}{{\em JHEP} {\bfseries 07}
		(2012) 088}, \href{http://arxiv.org/abs/1203.6620}{{\ttfamily arXiv:1203.6620
			[hep-th]}}.
	
	\bibitem{Cai:2012nm}
	R.-G. Cai, S.~He, L.~Li, and Y.-L. Zhang, ``{Holographic Entanglement Entropy
		on P-wave Superconductor Phase Transition},''
	\href{http://dx.doi.org/10.1007/JHEP07(2012)027}{{\em JHEP} {\bfseries 07}
		(2012) 027}, \href{http://arxiv.org/abs/1204.5962}{{\ttfamily arXiv:1204.5962
			[hep-th]}}.
	
	\bibitem{Jeong:2022zea}
	H.-S. Jeong, K.-Y. Kim, and Y.-W. Sun, ``{Holographic entanglement density for
		spontaneous symmetry breaking},''
	\href{http://dx.doi.org/10.1007/JHEP06(2022)078}{{\em JHEP} {\bfseries 06}
		(2022) 078}, \href{http://arxiv.org/abs/2203.07612}{{\ttfamily
			arXiv:2203.07612 [hep-th]}}.
	
	\bibitem{Yang:2023wuw}
	Z.~Yang, F.-J. Cheng, C.~Niu, C.-Y. Zhang, and P.~Liu, ``{The mixed-state
		entanglement in holographic p-wave superconductor model},''
	\href{http://dx.doi.org/10.1007/JHEP04(2023)110}{{\em JHEP} {\bfseries 04}
		(2023) 110}, \href{http://arxiv.org/abs/2301.13574}{{\ttfamily
			arXiv:2301.13574 [hep-th]}}.
	
	\bibitem{Klebanov:2007ws}
	I.~R. Klebanov, D.~Kutasov, and A.~Murugan, ``{Entanglement as a probe of
		confinement},'' \href{http://dx.doi.org/10.1016/j.nuclphysb.2007.12.017}{{\em
			Nucl. Phys. B} {\bfseries 796} (2008) 274--293},
	\href{http://arxiv.org/abs/0709.2140}{{\ttfamily arXiv:0709.2140 [hep-th]}}.
	
	\bibitem{Lewkowycz:2012mw}
	A.~Lewkowycz, ``{Holographic Entanglement Entropy and Confinement},''
	\href{http://dx.doi.org/10.1007/JHEP05(2012)032}{{\em JHEP} {\bfseries 05}
		(2012) 032}, \href{http://arxiv.org/abs/1204.0588}{{\ttfamily arXiv:1204.0588
			[hep-th]}}.
	
	\bibitem{Kol:2014nqa}
	U.~Kol, C.~Nunez, D.~Schofield, J.~Sonnenschein, and M.~Warschawski,
	``{Confinement, Phase Transitions and non-Locality in the Entanglement
		Entropy},'' \href{http://dx.doi.org/10.1007/JHEP06(2014)005}{{\em JHEP}
		{\bfseries 06} (2014) 005}, \href{http://arxiv.org/abs/1403.2721}{{\ttfamily
			arXiv:1403.2721 [hep-th]}}.
	
	\bibitem{Jain:2020rbb}
	P.~Jain and S.~Mahapatra, ``{Mixed state entanglement measures as probe for
		confinement},'' \href{http://dx.doi.org/10.1103/PhysRevD.102.126022}{{\em
			Phys. Rev. D} {\bfseries 102} (2020) 126022},
	\href{http://arxiv.org/abs/2010.07702}{{\ttfamily arXiv:2010.07702
			[hep-th]}}.
	
	\bibitem{Arefeva:2020uec}
	I.~Y. Aref'eva, A.~Patrushev, and P.~Slepov, ``{Holographic entanglement
		entropy in anisotropic background with confinement-deconfinement phase
		transition},'' \href{http://dx.doi.org/10.1007/JHEP07(2020)043}{{\em JHEP}
		{\bfseries 07} (2020) 043}, \href{http://arxiv.org/abs/2003.05847}{{\ttfamily
			arXiv:2003.05847 [hep-th]}}.
	
	\bibitem{Jain:2022hxl}
	P.~Jain, S.~S. Jena, and S.~Mahapatra, ``{Holographic confining-deconfining
		gauge theories and entanglement measures with a magnetic field},''
	\href{http://dx.doi.org/10.1103/PhysRevD.107.086016}{{\em Phys. Rev. D}
		{\bfseries 107} no.~8, (2023) 086016},
	\href{http://arxiv.org/abs/2209.15355}{{\ttfamily arXiv:2209.15355
			[hep-th]}}.
	
	\bibitem{daRocha:2021xwq}
	R.~da~Rocha, ``{Holographic entanglement entropy, deformed black branes, and
		deconfinement in AdS/QCD},''
	\href{http://dx.doi.org/10.1103/PhysRevD.105.026014}{{\em Phys. Rev. D}
		{\bfseries 105} no.~2, (2022) 026014},
	\href{http://arxiv.org/abs/2111.01244}{{\ttfamily arXiv:2111.01244
			[hep-th]}}.
	
	\bibitem{Chen:2023vjz}
	X.~Chen, B.~Yu, P.-C. Chu, and X.-H. Li, ``{The effect of gluon condensate on
		the entanglement entropy in a holographic model},''
	\href{http://arxiv.org/abs/2306.00682}{{\ttfamily arXiv:2306.00682
			[hep-ph]}}.
	
	\bibitem{Mahapatra:2019uql}
	S.~Mahapatra, ``{Interplay between the holographic QCD phase diagram and mutual
		\textbackslash{}\& $n$-partite information},''
	\href{http://dx.doi.org/10.1007/JHEP04(2019)137}{{\em JHEP} {\bfseries 04}
		(2019) 137}, \href{http://arxiv.org/abs/1903.05927}{{\ttfamily
			arXiv:1903.05927 [hep-th]}}.
	
	\bibitem{Ebrahim:2020qif}
	H.~Ebrahim and G.-M. Nafisi, ``{Holographic Mutual Information and Critical
		Exponents of the Strongly Coupled Plasma},''
	\href{http://dx.doi.org/10.1103/PhysRevD.102.106007}{{\em Phys. Rev. D}
		{\bfseries 102} no.~10, (2020) 106007},
	\href{http://arxiv.org/abs/2002.09993}{{\ttfamily arXiv:2002.09993
			[hep-th]}}.
	
	\bibitem{Walsh:2020lty}
	C.~Walsh, P.~S\'emon, D.~Poulin, G.~Sordi, and A.~M.~S. Tremblay,
	``{Entanglement and Classical Correlations at the Doping-Driven Mott
		Transition in the Two-Dimensional Hubbard Model},''
	\href{http://dx.doi.org/10.1103/PRXQuantum.1.020310}{{\em PRX Quantum}
		{\bfseries 1} no.~2, (2020) 020310}.
	
	\bibitem{Erlich:2005qh}
	J.~Erlich, E.~Katz, D.~T. Son, and M.~A. Stephanov, ``{QCD and a holographic
		model of hadrons},''
	\href{http://dx.doi.org/10.1103/PhysRevLett.95.261602}{{\em Phys. Rev. Lett.}
		{\bfseries 95} (2005) 261602},
	\href{http://arxiv.org/abs/hep-ph/0501128}{{\ttfamily arXiv:hep-ph/0501128}}.
	
	\bibitem{Karch:2006pv}
	A.~Karch, E.~Katz, D.~T. Son, and M.~A. Stephanov, ``{Linear confinement and
		AdS/QCD},'' \href{http://dx.doi.org/10.1103/PhysRevD.74.015005}{{\em Phys.
			Rev. D} {\bfseries 74} (2006) 015005},
	\href{http://arxiv.org/abs/hep-ph/0602229}{{\ttfamily arXiv:hep-ph/0602229}}.
	
	\bibitem{Hong:2007kx}
	D.~K. Hong, M.~Rho, H.-U. Yee, and P.~Yi, ``{Chiral Dynamics of Baryons from
		String Theory},'' \href{http://dx.doi.org/10.1103/PhysRevD.76.061901}{{\em
			Phys. Rev. D} {\bfseries 76} (2007) 061901},
	\href{http://arxiv.org/abs/hep-th/0701276}{{\ttfamily arXiv:hep-th/0701276}}.
	
	\bibitem{Nawa:2006gv}
	K.~Nawa, H.~Suganuma, and T.~Kojo, ``{Baryons in holographic QCD},''
	\href{http://dx.doi.org/10.1103/PhysRevD.75.086003}{{\em Phys. Rev. D}
		{\bfseries 75} (2007) 086003},
	\href{http://arxiv.org/abs/hep-th/0612187}{{\ttfamily arXiv:hep-th/0612187}}.
	
	\bibitem{Colangelo:2007pt}
	P.~Colangelo, F.~De~Fazio, F.~Jugeau, and S.~Nicotri, ``{On the light glueball
		spectrum in a holographic description of QCD},''
	\href{http://dx.doi.org/10.1016/j.physletb.2007.06.072}{{\em Phys. Lett. B}
		{\bfseries 652} (2007) 73--78},
	\href{http://arxiv.org/abs/hep-ph/0703316}{{\ttfamily arXiv:hep-ph/0703316}}.
	
	\bibitem{Boschi-Filho:2005xct}
	H.~Boschi-Filho, N.~R.~F. Braga, and H.~L. Carrion, ``{Glueball Regge
		trajectories from gauge/string duality and the Pomeron},''
	\href{http://dx.doi.org/10.1103/PhysRevD.73.047901}{{\em Phys. Rev. D}
		{\bfseries 73} (2006) 047901},
	\href{http://arxiv.org/abs/hep-th/0507063}{{\ttfamily arXiv:hep-th/0507063}}.
	
	\bibitem{Li:2011hp}
	D.~Li, S.~He, M.~Huang, and Q.-S. Yan, ``{Thermodynamics of deformed AdS$_5$
		model with a positive/negative quadratic correction in graviton-dilaton
		system},'' \href{http://dx.doi.org/10.1007/JHEP09(2011)041}{{\em JHEP}
		{\bfseries 09} (2011) 041}, \href{http://arxiv.org/abs/1103.5389}{{\ttfamily
			arXiv:1103.5389 [hep-th]}}.
	
	\bibitem{Li:2012ay}
	D.~Li, M.~Huang, and Q.-S. Yan, ``{A dynamical soft-wall holographic QCD model
		for chiral symmetry breaking and linear confinement},''
	\href{http://dx.doi.org/10.1140/epjc/s10052-013-2615-3}{{\em Eur. Phys. J. C}
		{\bfseries 73} (2013) 2615}, \href{http://arxiv.org/abs/1206.2824}{{\ttfamily
			arXiv:1206.2824 [hep-th]}}.
	
	\bibitem{Fang:2015ytf}
	Z.~Fang, S.~He, and D.~Li, ``{Chiral and Deconfining Phase Transitions from
		Holographic QCD Study},''
	\href{http://dx.doi.org/10.1016/j.nuclphysb.2016.04.003}{{\em Nucl. Phys. B}
		{\bfseries 907} (2016) 187--207},
	\href{http://arxiv.org/abs/1512.04062}{{\ttfamily arXiv:1512.04062
			[hep-ph]}}.
	
	\bibitem{Fang:2019lsz}
	Z.~Fang and Y.-L. Wu, ``{Equation of state and chiral transition in soft-wall
		AdS/QCD with more realistic gravitational background},''
	\href{http://arxiv.org/abs/1909.06917}{{\ttfamily arXiv:1909.06917
			[hep-ph]}}.
	
	\bibitem{Li:2013oda}
	D.~Li and M.~Huang, ``{Dynamical holographic QCD model for glueball and light
		meson spectra},'' \href{http://dx.doi.org/10.1007/JHEP11(2013)088}{{\em JHEP}
		{\bfseries 11} (2013) 088}, \href{http://arxiv.org/abs/1303.6929}{{\ttfamily
			arXiv:1303.6929 [hep-ph]}}.
	
	\bibitem{Gutsche:2019pls}
	T.~Gutsche, V.~E. Lyubovitskij, I.~Schmidt, and A.~Y. Trifonov, ``{Baryons in a
		soft-wall AdS-Schwarzschild approach at low temperature},''
	\href{http://dx.doi.org/10.1103/PhysRevD.99.114023}{{\em Phys. Rev. D}
		{\bfseries 99} no.~11, (2019) 114023},
	\href{http://arxiv.org/abs/1905.02577}{{\ttfamily arXiv:1905.02577
			[hep-ph]}}.
	
	\bibitem{Gutsche:2019blp}
	T.~Gutsche, V.~E. Lyubovitskij, I.~Schmidt, and A.~Y. Trifonov, ``{Mesons in a
		soft-wall AdS-Schwarzschild approach at low temperature},''
	\href{http://dx.doi.org/10.1103/PhysRevD.99.054030}{{\em Phys. Rev. D}
		{\bfseries 99} no.~5, (2019) 054030},
	\href{http://arxiv.org/abs/1902.01312}{{\ttfamily arXiv:1902.01312
			[hep-ph]}}.
	
	\bibitem{Li:2014dsa}
	D.~Li, S.~He, and M.~Huang, ``{Temperature dependent transport coefficients in
		a dynamical holographic QCD model},''
	\href{http://dx.doi.org/10.1007/JHEP06(2015)046}{{\em JHEP} {\bfseries 06}
		(2015) 046}, \href{http://arxiv.org/abs/1411.5332}{{\ttfamily arXiv:1411.5332
			[hep-ph]}}.
	
	\bibitem{Gutsche:2017lyu}
	T.~Gutsche, V.~E. Lyubovitskij, and I.~Schmidt, ``{Electromagnetic structure of
		nucleon and Roper in soft-wall AdS/QCD},''
	\href{http://dx.doi.org/10.1103/PhysRevD.97.054011}{{\em Phys. Rev. D}
		{\bfseries 97} no.~5, (2018) 054011},
	\href{http://arxiv.org/abs/1712.08410}{{\ttfamily arXiv:1712.08410
			[hep-ph]}}.
	
	\bibitem{Gutsche:2019jzh}
	T.~Gutsche, V.~E. Lyubovitskij, and I.~Schmidt, ``{Electromagnetic properties
		of the nucleon and the Roper resonance in soft-wall AdS/QCD at finite
		temperature},'' \href{http://dx.doi.org/10.1016/j.nuclphysb.2020.114934}{{\em
			Nucl. Phys. B} {\bfseries 952} (2020) 114934},
	\href{http://arxiv.org/abs/1906.08641}{{\ttfamily arXiv:1906.08641
			[hep-ph]}}.
	
	\bibitem{DeWolfe:2010he}
	O.~DeWolfe, S.~S. Gubser, and C.~Rosen, ``{A holographic critical point},''
	\href{http://dx.doi.org/10.1103/PhysRevD.83.086005}{{\em Phys. Rev. D}
		{\bfseries 83} (2011) 086005},
	\href{http://arxiv.org/abs/1012.1864}{{\ttfamily arXiv:1012.1864 [hep-th]}}.
	
	\bibitem{DeWolfe:2011ts}
	O.~DeWolfe, S.~S. Gubser, and C.~Rosen, ``{Dynamic critical phenomena at a
		holographic critical point},''
	\href{http://dx.doi.org/10.1103/PhysRevD.84.126014}{{\em Phys. Rev. D}
		{\bfseries 84} (2011) 126014},
	\href{http://arxiv.org/abs/1108.2029}{{\ttfamily arXiv:1108.2029 [hep-th]}}.
	
	\bibitem{Cai:2012xh}
	R.-G. Cai, S.~He, and D.~Li, ``{A hQCD model and its phase diagram in
		Einstein-Maxwell-Dilaton system},''
	\href{http://dx.doi.org/10.1007/JHEP03(2012)033}{{\em JHEP} {\bfseries 03}
		(2012) 033}, \href{http://arxiv.org/abs/1201.0820}{{\ttfamily arXiv:1201.0820
			[hep-th]}}.
	
	\bibitem{Cai:2012eh}
	R.-G. Cai, S.~Chakrabortty, S.~He, and L.~Li, ``{Some aspects of QGP phase in a
		hQCD model},'' \href{http://dx.doi.org/10.1007/JHEP02(2013)068}{{\em JHEP}
		{\bfseries 02} (2013) 068}, \href{http://arxiv.org/abs/1209.4512}{{\ttfamily
			arXiv:1209.4512 [hep-th]}}.
	
	\bibitem{Finazzo:2013efa}
	S.~I. Finazzo and J.~Noronha, ``{Holographic calculation of the electric
		conductivity of the strongly coupled quark-gluon plasma near the
		deconfinement transition},''
	\href{http://dx.doi.org/10.1103/PhysRevD.89.106008}{{\em Phys. Rev. D}
		{\bfseries 89} no.~10, (2014) 106008},
	\href{http://arxiv.org/abs/1311.6675}{{\ttfamily arXiv:1311.6675 [hep-th]}}.
	
	\bibitem{Yang:2014bqa}
	Y.~Yang and P.-H. Yuan, ``{A Refined Holographic QCD Model and QCD Phase
		Structure},'' \href{http://dx.doi.org/10.1007/JHEP11(2014)149}{{\em JHEP}
		{\bfseries 11} (2014) 149}, \href{http://arxiv.org/abs/1406.1865}{{\ttfamily
			arXiv:1406.1865 [hep-th]}}.
	
	\bibitem{Critelli:2017oub}
	R.~Critelli, J.~Noronha, J.~Noronha-Hostler, I.~Portillo, C.~Ratti, and
	R.~Rougemont, ``{Critical point in the phase diagram of primordial
		quark-gluon matter from black hole physics},''
	\href{http://dx.doi.org/10.1103/PhysRevD.96.096026}{{\em Phys. Rev. D}
		{\bfseries 96} no.~9, (2017) 096026},
	\href{http://arxiv.org/abs/1706.00455}{{\ttfamily arXiv:1706.00455
			[nucl-th]}}.
	
	\bibitem{Li:2017ple}
	Z.~Li, Y.~Chen, D.~Li, and M.~Huang, ``{Locating the QCD critical end point
		through the peaked baryon number susceptibilities along the freeze-out
		line},'' \href{http://dx.doi.org/10.1088/1674-1137/42/1/013103}{{\em Chin.
			Phys. C} {\bfseries 42} no.~1, (2018) 013103},
	\href{http://arxiv.org/abs/1706.02238}{{\ttfamily arXiv:1706.02238
			[hep-ph]}}.
	
	\bibitem{Chen:2017cyc}
	Y.~Chen, M.~Huang, and Q.-S. Yan, ``{Gravitation waves from QCD and electroweak
		phase transitions},'' \href{http://dx.doi.org/10.1007/JHEP05(2018)178}{{\em
			JHEP} {\bfseries 05} (2018) 178},
	\href{http://arxiv.org/abs/1712.03470}{{\ttfamily arXiv:1712.03470
			[hep-ph]}}.
	
	\bibitem{Knaute:2017opk}
	J.~Knaute, R.~Yaresko, and B.~K\"ampfer, ``{Holographic QCD phase diagram with
		critical point from Einstein\textendash{}Maxwell-dilaton dynamics},''
	\href{http://dx.doi.org/10.1016/j.physletb.2018.01.053}{{\em Phys. Lett. B}
		{\bfseries 778} (2018) 419--425},
	\href{http://arxiv.org/abs/1702.06731}{{\ttfamily arXiv:1702.06731
			[hep-ph]}}.
	
	\bibitem{Fang:2018axm}
	Z.~Fang, Y.-L. Wu, and L.~Zhang, ``{Chiral phase transition and QCD phase
		diagram from AdS/QCD},''
	\href{http://dx.doi.org/10.1103/PhysRevD.99.034028}{{\em Phys. Rev. D}
		{\bfseries 99} no.~3, (2019) 034028},
	\href{http://arxiv.org/abs/1810.12525}{{\ttfamily arXiv:1810.12525
			[hep-ph]}}.
	
	\bibitem{Ballon-Bayona:2020xls}
	A.~Ballon-Bayona, H.~Boschi-Filho, E.~F. Capossoli, and D.~M. Rodrigues,
	``{Criticality from Einstein-Maxwell-dilaton holography at finite temperature
		and density},'' \href{http://dx.doi.org/10.1103/PhysRevD.102.126003}{{\em
			Phys. Rev. D} {\bfseries 102} no.~12, (2020) 126003},
	\href{http://arxiv.org/abs/2006.08810}{{\ttfamily arXiv:2006.08810
			[hep-th]}}.
	
	\bibitem{Li:2020hau}
	M.-W. Li, Y.~Yang, and P.-H. Yuan, ``{Analytic Study on Chiral Phase Transition
		in Holographic QCD},'' \href{http://dx.doi.org/10.1007/JHEP02(2021)055}{{\em
			JHEP} {\bfseries 02} (2021) 055},
	\href{http://arxiv.org/abs/2009.05694}{{\ttfamily arXiv:2009.05694
			[hep-th]}}.
	
	\bibitem{Grefa:2021qvt}
	J.~Grefa, J.~Noronha, J.~Noronha-Hostler, I.~Portillo, C.~Ratti, and
	R.~Rougemont, ``{Hot and dense quark-gluon plasma thermodynamics from
		holographic black holes},''
	\href{http://dx.doi.org/10.1103/PhysRevD.104.034002}{{\em Phys. Rev. D}
		{\bfseries 104} no.~3, (2021) 034002},
	\href{http://arxiv.org/abs/2102.12042}{{\ttfamily arXiv:2102.12042
			[nucl-th]}}.
	
	\bibitem{He:2022amv}
	S.~He, L.~Li, Z.~Li, and S.-J. Wang, ``{Gravitational Waves and Primordial
		Black Hole Productions from Gluodynamics},''
	\href{http://arxiv.org/abs/2210.14094}{{\ttfamily arXiv:2210.14094
			[hep-ph]}}.
	
	\bibitem{Grefa:2022fpu}
	J.~Grefa, M.~Hippert, J.~Noronha, J.~Noronha-Hostler, I.~Portillo, C.~Ratti,
	and R.~Rougemont, ``{QCD Equilibrium and Dynamical Properties from
		Holographic Black Holes},''
	\href{http://dx.doi.org/10.31349/SuplRevMexFis.3.040910}{{\em Rev. Mex. Fis.
			Suppl.} {\bfseries 3} no.~4, (2022) 040910},
	\href{http://arxiv.org/abs/2207.12564}{{\ttfamily arXiv:2207.12564
			[nucl-th]}}.
	
	\bibitem{Chen:2024ckb}
	X.~Chen and M.~Huang, ``{Machine learning holographic black hole from lattice
		QCD equation of state},''
	\href{http://dx.doi.org/10.1103/PhysRevD.109.L051902}{{\em Phys. Rev. D}
		{\bfseries 109} no.~5, (2024) L051902},
	\href{http://arxiv.org/abs/2401.06417}{{\ttfamily arXiv:2401.06417
			[hep-ph]}}.
	
	\bibitem{Jokela:2023lvr}
	N.~Jokela, H.~Ruotsalainen, and J.~G. Subils, ``{Limitations of entanglement
		entropy in detecting thermal phase transitions},''
	\href{http://dx.doi.org/10.1007/JHEP01(2024)186}{{\em JHEP} {\bfseries 01}
		(2024) 186}, \href{http://arxiv.org/abs/2310.11205}{{\ttfamily
			arXiv:2310.11205 [hep-th]}}.
	
	\bibitem{Cai:2022omk}
	R.-G. Cai, S.~He, L.~Li, and Y.-X. Wang, ``{Probing QCD critical point and
		induced gravitational wave by black hole physics},''
	\href{http://dx.doi.org/10.1103/PhysRevD.106.L121902}{{\em Phys. Rev. D}
		{\bfseries 106} no.~12, (2022) L121902},
	\href{http://arxiv.org/abs/2201.02004}{{\ttfamily arXiv:2201.02004
			[hep-th]}}.
	
	\bibitem{HotQCD:2012fhj}
	{\bfseries HotQCD} Collaboration, A.~Bazavov {\em et~al.}, ``{Fluctuations and
		Correlations of net baryon number, electric charge, and strangeness: A
		comparison of lattice QCD results with the hadron resonance gas model},''
	\href{http://dx.doi.org/10.1103/PhysRevD.86.034509}{{\em Phys. Rev. D}
		{\bfseries 86} (2012) 034509},
	\href{http://arxiv.org/abs/1203.0784}{{\ttfamily arXiv:1203.0784 [hep-lat]}}.
	
	\bibitem{Bazavov:2017dus}
	A.~Bazavov {\em et~al.}, ``{The QCD Equation of State to $\mathcal{O}(\mu_B^6)$
		from Lattice QCD},'' \href{http://dx.doi.org/10.1103/PhysRevD.95.054504}{{\em
			Phys. Rev. D} {\bfseries 95} no.~5, (2017) 054504},
	\href{http://arxiv.org/abs/1701.04325}{{\ttfamily arXiv:1701.04325
			[hep-lat]}}.
	
	\bibitem{Zhao:2022uxc}
	Y.-Q. Zhao, S.~He, D.~Hou, L.~Li, and Z.~Li, ``{Phase diagram of holographic
		thermal dense QCD matter with rotation},''
	\href{http://dx.doi.org/10.1007/JHEP04(2023)115}{{\em JHEP} {\bfseries 04}
		(2023) 115}, \href{http://arxiv.org/abs/2212.14662}{{\ttfamily
			arXiv:2212.14662 [hep-ph]}}.
	
	\bibitem{Li:2023mpv}
	Z.~Li, J.~Liang, S.~He, and L.~Li, ``{Holographic study of higher-order baryon
		number susceptibilities at finite temperature and density},''
	\href{http://dx.doi.org/10.1103/PhysRevD.108.046008}{{\em Phys. Rev. D}
		{\bfseries 108} no.~4, (2023) 046008},
	\href{http://arxiv.org/abs/2305.13874}{{\ttfamily arXiv:2305.13874
			[hep-ph]}}.
	
	\bibitem{Rougemont:2023gfz}
	R.~Rougemont, J.~Grefa, M.~Hippert, J.~Noronha, J.~Noronha-Hostler,
	I.~Portillo, and C.~Ratti, ``{Hot QCD Phase Diagram From Holographic
		Einstein-Maxwell-Dilaton Models},''
	\href{http://arxiv.org/abs/2307.03885}{{\ttfamily arXiv:2307.03885
			[nucl-th]}}.
	
	\bibitem{Lewkowycz:2013nqa}
	A.~Lewkowycz and J.~Maldacena, ``{Generalized gravitational entropy},''
	\href{http://dx.doi.org/10.1007/JHEP08(2013)090}{{\em JHEP} {\bfseries 08}
		(2013) 090}, \href{http://arxiv.org/abs/1304.4926}{{\ttfamily arXiv:1304.4926
			[hep-th]}}.
	
	\bibitem{Huang:2014aga}
	W.-H. Huang, ``{Generalized Gravitational Entropy of Interacting Scalar Field
		and Maxwell Field},''
	\href{http://dx.doi.org/10.1016/j.physletb.2014.11.012}{{\em Phys. Lett. B}
		{\bfseries 739} (2014) 365--369},
	\href{http://arxiv.org/abs/1409.4893}{{\ttfamily arXiv:1409.4893 [hep-th]}}.
	
	\bibitem{Li:2016pwu}
	Z.~Li and J.-j. Zhang, ``{On one-loop entanglement entropy of two short
		intervals from OPE of twist operators},''
	\href{http://dx.doi.org/10.1007/JHEP05(2016)130}{{\em JHEP} {\bfseries 05}
		(2016) 130}, \href{http://arxiv.org/abs/1604.02779}{{\ttfamily
			arXiv:1604.02779 [hep-th]}}.
	
	\bibitem{Chen:2017ahf}
	B.~Chen, Z.~Li, and J.-j. Zhang, ``{Corrections to holographic entanglement
		plateau},'' \href{http://dx.doi.org/10.1007/JHEP09(2017)151}{{\em JHEP}
		{\bfseries 09} (2017) 151}, \href{http://arxiv.org/abs/1707.07354}{{\ttfamily
			arXiv:1707.07354 [hep-th]}}.
	
	\bibitem{Shuryak:2014zxa}
	E.~Shuryak, ``{Strongly coupled quark-gluon plasma in heavy ion collisions},''
	\href{http://dx.doi.org/10.1103/RevModPhys.89.035001}{{\em Rev. Mod. Phys.}
		{\bfseries 89} (2017) 035001},
	\href{http://arxiv.org/abs/1412.8393}{{\ttfamily arXiv:1412.8393 [hep-ph]}}.
	
	\bibitem{Ghiglieri:2020dpq}
	J.~Ghiglieri, A.~Kurkela, M.~Strickland, and A.~Vuorinen, ``{Perturbative
		Thermal QCD: Formalism and Applications},''
	\href{http://dx.doi.org/10.1016/j.physrep.2020.07.004}{{\em Phys. Rept.}
		{\bfseries 880} (2020) 1--73},
	\href{http://arxiv.org/abs/2002.10188}{{\ttfamily arXiv:2002.10188
			[hep-ph]}}.
	
\end{thebibliography}

\providecommand{\href}[2]{#2}\begingroup\raggedright\endgroup

\end{document}